\documentclass[
reprint,
superscriptaddress,
showpacs,
preprintnumbers,
nofootinbib,
nobibnotes,
amsmath,
amssymb, 
aps,
prd,
floatfix
]{revtex4-1}

\usepackage[utf8]{inputenc}
\usepackage[normalem]{ulem}
\usepackage{graphicx}
\usepackage{dcolumn}
\usepackage[colorlinks=true,allcolors=purple]{hyperref}
\usepackage{url}
\usepackage{enumerate}

\usepackage{mathrsfs} 
\usepackage[dvipsnames,table]{xcolor}
\usepackage{cancel}
\usepackage{bbold}
\usepackage{braket}
\usepackage{physics}
\usepackage{multirow}
\usepackage[capitalize]{cleveref}
\usepackage{xspace}
\usepackage{comment}
\usepackage{fontawesome} 
\definecolor{nicegreen}{rgb}{0., 0.75, 0.46}

\definecolor{wildstrawberry}{rgb}{1.0, 0.26, 0.64}

\definecolor{MH}{rgb}{0.0,0.6,9}
\definecolor{palatinate}{rgb}{0.494, 0.192, 0.482}
\definecolor{blue-violet}{rgb}{0.33, 0.17, 0.89}

\renewcommand{\phi}{\varphi}
\def\ltap{\ \raise.3ex\hbox{$<$\kern-.75em\lower1ex\hbox{$\sim$}}\ }
\def\gtap{\ \raise.3ex\hbox{$>$\kern-.75em\lower1ex\hbox{$\sim$}}\ }
\def\lsim{\ \raise.3ex\hbox{$<$\kern-.75em\lower1ex\hbox{$\sim$}}\ }
\def\gsim{\ \raise.3ex\hbox{$>$\kern-.75em\lower1ex\hbox{$\sim$}}\ }

\newcommand{\be}{\begin{equation}}
\newcommand{\ee}{\end{equation}}
\newcommand{\beq}{\begin{equation}}
\newcommand{\eeq}{\end{equation}}
\newcommand{\bea}{\begin{eqnarray}}
\newcommand{\eea}{\end{eqnarray}}
\newcommand{\bear}{\begin{eqnarray}}
\newcommand{\eear}{\end{eqnarray}}

\def\GeV{\,{\rm GeV}}

\def\MeV{\,{\rm MeV}}

\begin{document}
\preprint{MI-HET-843}
\preprint{CALT-TH/2024-036}
\preprint{FERMILAB-PUB-24-0869-T}
\title{Long-lived vectors from electromagnetic cascades at SHiP}

\author{Tao Zhou}
\email{taozhou@tamu.edu}
\affiliation{Department of Physics and Astronomy, Mitchell Institute for Fundamental Physics and Astronomy, Texas A\&M University, College Station, TX 77843, USA}
\author{Ryan Plestid}
\email{rplestid@caltech.edu}
\affiliation{Walter Burke Institute for Theoretical Physics, California Institute of Technology, Pasadena, CA 91125, USA}
\author{Kevin J. Kelly}
\email{kjkelly@tamu.edu}
\affiliation{Department of Physics and Astronomy, Mitchell Institute for Fundamental Physics and Astronomy, Texas A\&M University, College Station, TX 77843, USA}
\author{Nikita Blinov}
\email{nblinov@yorku.ca}
\affiliation{Department of Physics and Astronomy, York University, Toronto, Ontario, M3J 1P3, Canada}
\author{Patrick J. Fox}
\email{pjfox@fnal.gov}
\affiliation{Particle Theory Department, Fermi National Accelerator Laboratory, Batavia, IL 60510, USA}

\date{\today}

\begin{abstract}
    We simulate dark-vector, $V$, production from electromagnetic cascades at the recently approved SHiP experiment. The cascades (initiated by photons from $\pi^0\rightarrow \gamma \gamma$) can lead to 3-4 orders of magnitude increase of the event rate relative to using primary production alone. We provide new SHiP sensitivity projections for dark photons and electrophilic gauge bosons, which are significantly improved compared to previous literature. The main gain in sensitivity occurs for long-lived dark vectors with masses below $\sim 50-300~{\rm MeV}$. The dominant production mode in this parameter space is low-energy annihilation $e^+ e^- \rightarrow V(\gamma)$. This motivates a detailed study of backgrounds and efficiencies in the SHiP experiment for sub-GeV signals. 
\end{abstract}

\maketitle

\section{Introduction \label{sec:Introduction} }

Fixed-target experiments are indispensable tools when searching for rare events \cite{Artuso:2022ouk} and/or feebly interacting particles \cite{Antel:2023hkf}. Such beyond Standard Model (SM) particles are motivated by models of light dark matter, the origin of neutrino masses, and the strong-CP problem \cite{Antel:2023hkf}. 
A broad, competitive, and comprehensive experimental program has emerged with relevant facilities ranging from neutrino experiments at spallation sources \cite{Baxter:2019mcx,Barbeau:2021exu,JSNS2:2021hyk,CCM:2021leg,Aguilar-Arevalo:2023dai}, KEK \cite{T2K:2011qtm}, and Fermilab \cite{Machado:2019oxb,ArgoNeuT:2022mrm,DUNE:2022aul}, to electron beam dumps at CERN \cite{Banerjee:2019pds}, SLAC \cite{LDMX:2018cma}, and J-LAB \cite{Battaglieri:2022dcy}. In this work we focus on the Search for Hidden Particles (SHiP) experiment at CERN's SPS, which was approved in March 2024~\cite{SHiP:2021nfo,Aberle:2839677,Albanese:2878604}. For a comprehensive list of other current and planned dark sector experiments see~\cite{Beacham:2019nyx,Ilten:2022lfq}.

All of the fixed-target facilities use a high intensity beam with energies ranging from ${\sim}1~{\rm GeV}$ to $400~{\rm GeV}$. 
While some search strategies rely on missing energy and/or momentum, a powerful search technique is to look for visible signatures in a downstream detector, for example from the decay of a dark sector particle into SM particles. When pursuing this strategy, experimental sensitivities (plotted in the plane of coupling vs.~mass) typically have a ``ceiling'' and a ``floor'' (see  Ref.~\cite{Beacham:2019nyx} for examples) -- the ceiling is set by the decay length of the dark sector particle becoming much shorter than the thickness of the shielding before the decay volume/detector; the floor is set by the rarity of production, and by the shrinking probability of detection (either via decay or scattering) as the lifetime of the dark state grows. We focus on the case of visible decays e.g.,~$V\rightarrow e^+e^-$, in a decay pipe (e.g., at SHiP) and will refer to the floor of sensitivity as the ``lifetime frontier.''

In this paper we study the production of new particles in an electromagnetic cascade initiated by photons (themselves produced via meson decays e.g., $\pi^0 \rightarrow \gamma \gamma$). As discussed in Refs.~\cite{Donnelly:1978ty,Tsai:1986tx,Bjorken:1988as,Prinz:1998ua,Prinz:2001qz,Marsicano:2018glj,Marsicano:2018glj,Nardi:2018cxi,Marsicano:2018krp,Celentano:2020vtu,Dutta:2020vop,Brdar:2020dpr,Capozzi:2021nmp,Andreev:2021fzd,Brdar:2022vum,Darme:2022zfw,NA64:2023wbi,Blinov:2024gcw,Arias-Aragon:2024qji,Arias-Aragon:2024gpm}, accounting for electromagnetic cascades can substantially enhance reach for sub-GeV mass long-lived particles (i.e.,~beyond that estimated using only primary production \cite{Bauer:2018onh,SHiP:2020vbd}). The improvement in sensitivity can be expressed as gain per $\pi^0$, which is relatively insensitive to the hadronic modelling used to prepare a $\pi^0$ sample.

Electromagnetic cascades give rise to a large multiplicity of increasingly soft photons, electrons, and positrons all of which can themselves produce feebly interacting particles. We use the Package for Electromagnetic Transitions In Thick-target Environments (\texttt{PETITE} {\href{https://github.com/kjkellyphys/PETITE}{\large\color{BlueViolet}\faGithub}}~\cite{Blinov:2024gcw}) to simulate both the SM electromagnetic cascade and the production of particles beyond the SM. The low energy of the parent particles results in a flux of dark vectors which has a lower average energy, and therefore shorter average decay length, than the flux produced by high energy processes (such as the decay of highly boosted mesons). This feature, coupled with the growing multiplicity of parents in the shower, leads to substantial gains in sensitivity at the lifetime frontier. 

As an illustrative example, we study dark vectors, $V$, coupled to electron's vector current, 
\begin{equation}
    \mathscr{L} \subset -\frac14 \mathcal{F}_{\mu\nu} \mathcal{F}^{\mu\nu} +\frac{1}{2}m_V^2 V_\mu V^\mu + g_V \bar{e} \gamma_\mu e V^\mu~,
\end{equation}
where $V$ is a vector-field with field strength $\mathcal{F}$ and mass $m_V$. 
We focus on regions of parameter space ($m_V\lesssim 1~{\rm GeV}$ and $10^{-9}\lesssim g_{V} \lesssim 10^{-3}$) where the SHiP experiment will have good sensitivity. Dark vectors arise as gauge bosons in extensions of the SM where, e.g., a flavor subgroup has been gauged (such as $B-L$ or $L_e-L_\tau$), or where a new $U(1)$ (i.e., a dark photon) is introduced. In the former case, 
the boson couples directly to the corresponding current via the gauge coupling $g_V$. For the dark photon the coupling to matter is through a kinetic mixing with the photon, leading to interactions with the electromagnetic current and $g_V=\varepsilon e$, where $e$ is the charge of an electron, and $\varepsilon$ is the loop-induced mixing parameter~\cite{Holdom:1985ag}. In both cases the gauge boson obtains a mass either through spontaneous symmetry breaking, or through the Stueckelberg mechanism~\cite{Stueckelberg:1938hvi}.

As three relevant examples we choose to focus on the dark photon and the two anomaly-free electrophilic models: $L_e - L_\mu$ and $L_e - L_\tau$~\cite{He:1991qd,Araki:2012ip,Bauer:2018onh}. We have also considered $L_\mu-L_\tau$, however we find that in this case the electromagnetic cascade does not dominate sensitivities in the same manner as for the dark photon and the electrophilic $L_e-L_{\mu,\tau}$ models. We defer a detailed analysis of $L_\mu-L_\tau$ to future work, but briefly comment on the main qualitative features in \cref{sec:Lmu-Ltau}. 
These choices cover a broad range of different behaviors, in which the various electromagnetic processes benefit (relatively) from the electrophilic coupling in the $L_e - L_{\mu,\tau}$ scenarios, in contrast with the democratically coupled dark photon. Although $L_e-L_\mu$ and $L_e-L_\tau$ have qualitatively similar phenomenology, we include both for completeness. 

Weighting events by their probability of decay in the decay volume of the detector, the dominant production mechanism over large regions of parameter space turns out to be  $e^+e^- \rightarrow V(\gamma)$, i.e., resonant annihilation of positrons colliding with atomic electrons in the fixed target. The results of this paper compliment recent studies of positron annihilation in the context of new physics searches at PADME \cite{Nardi:2018cxi,Marsicano:2018krp,Marsicano:2018glj,Darme:2022zfw,Arias-Aragon:2024qji}, neutrino experiments \cite{Celentano:2020vtu,Capozzi:2021nmp,Dutta:2020vop,Brdar:2020dpr,Brdar:2022vum,Capozzi:2023ffu}, and fixed target experiments more generally \cite{Arias-Aragon:2024gpm}. We find that including electromagnetic cascades can increase the reach of SHiP in $g_V$ by as over an order of magnitude for low vector masses.

The rest of the paper is organized along the following lines. In \cref{sec:Production} we review the dark vector production mechanisms in proton beam dumps. We focus primarily on direct production from meson decay, and subsequent production from the electromagnetic cascade. Special attention is paid to resonant annihilation of positrons. Next, in \cref{sec:Lifetime} we explain how different production mechanisms dominate sensitivity at different values of the coupling $g_V$. As $g_V\rightarrow 0$, decay lengths become much longer than the size of the experiment, and the importance of the low-energy flux is enhanced by the larger probability of decay within the decay pipe. 
\cref{sec:Sensitivity} provides updated SHiP sensitivity projections and discusses how each production mechanism contributes to the overall sensitivity. Finally, in \cref{sec:Conclusions} we comment on implications for SHiP's search strategy and outline potentially interesting avenues for future investigation.

\begin{figure*}[t]
\centering
\includegraphics[width=0.80\linewidth]{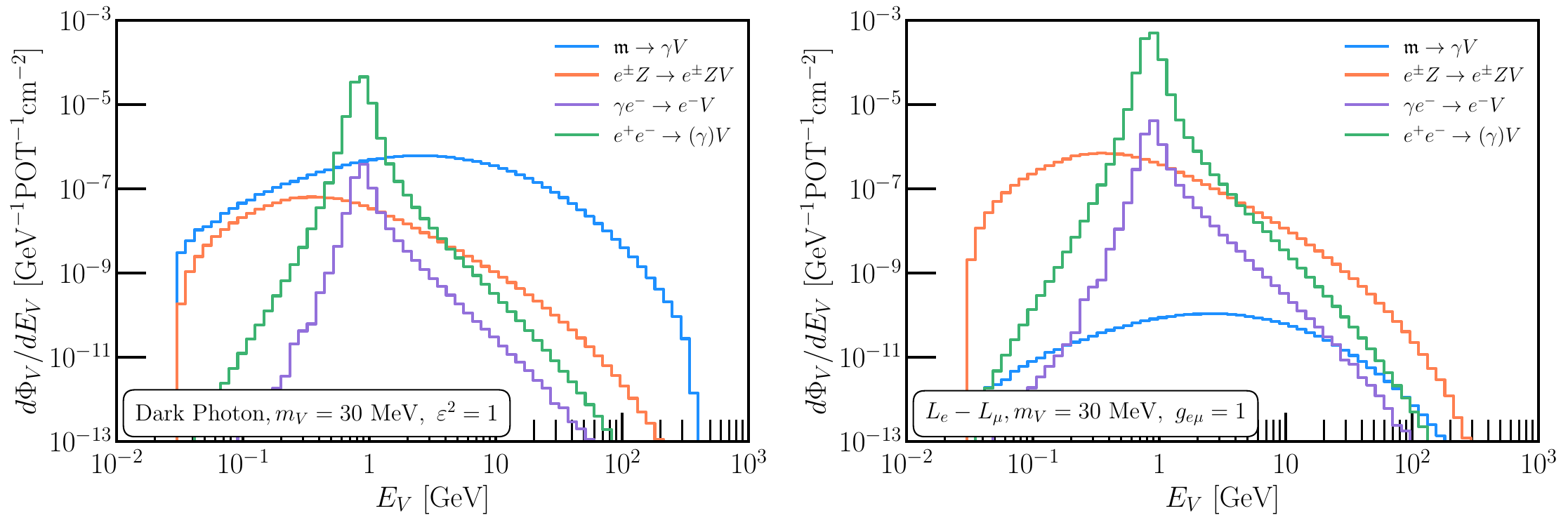}\\
\includegraphics[width=0.80\linewidth]{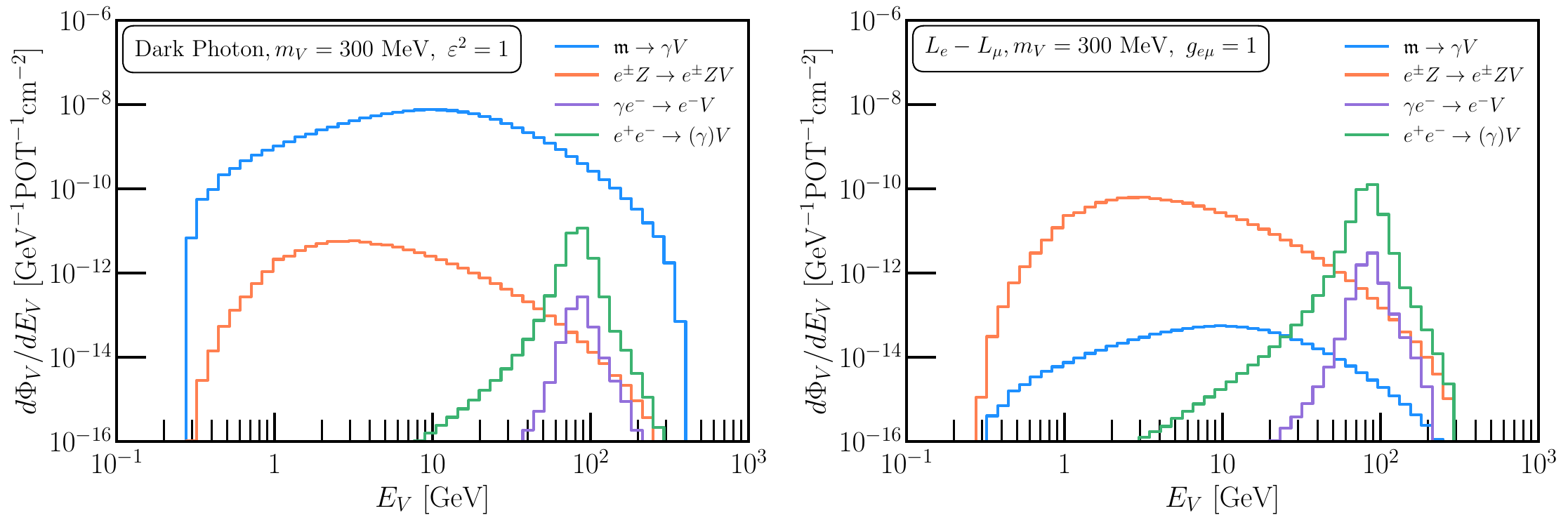}
\caption{Top row: Expected flux of $m_V = 30$ MeV dark vectors in the dark photon (left) and $L_e - L_\mu$ gauge boson (right) models at SHiP, assuming produced particles are stable. Each color (blue: $\mathfrak{m}$ for meson ($\pi^0$, $\eta^{(\prime)}$, $\omega$) decay, orange: $e^\pm$ bremsstrahlung, purple: Compton upscattering, green: positron annihilation) represents a different production mechanism, as discussed in the text, in the molybdenum target of SHiP. In the right panel, we assume that the kinetic mixing $\varepsilon^2$ is only generated at loop level from SM lepton contributions. As a result, the $V$ flux from meson decays is suppressed.
Bottom row: same as above but for $m_V = 300$ MeV.
\label{fig:energy-spectrum}}
\end{figure*}

\section{Production Mechanisms}\label{sec:Production}

The SHiP experiment will collide the CERN SPS 400~GeV proton beam into a molybdenum (${\rm Mo}$ with $Z=42$) and tungsten ($W$ with $Z=74$) target. The first $\sim 60$ radiation lengths (580~{\rm mm} as shown in Table II of Ref.~\cite{LopezSola:2019sfp}) are primarily molybdenum with sub-percent contributions from tungsten; the bulk of the electromagnetic cascades occur in the molybdenum and so we assume a uniform molybdenum target in our simulations for simplicity. 
This beam energy corresponds to a  proton-proton center of mass energy $\sqrt{s_{pp}} \gtrsim 25~{\rm GeV}$ so that a wide range of final states is possible. Well-studied production mechanisms in the literature include Drell-Yan like processes, proton-nucleon bremsstrahlung, and production from the decay of neutral mesons that are produced in $pp$ collisions \cite{deNiverville:2011it,SHiP:2015vad,Bauer:2018onh,SHiP:2020vbd}. Since meson production dominates sensitivity in a wide range of parameter space, and we will focus on a photon flux produced by the decay of mesons, we neglect Drell-Yan and proton bremsstrahlung as production modes in what follows; they can be added to what is presented below and are important in complementary regions of parameter space. 

In~\cref{fig:energy-spectrum} we plot the flux of simulated dark vectors with masses $m_V=30~{\rm MeV}$ and $300~{\rm MeV}$ as a function of $E_V$.\!\footnote{Specifically we display the flux at the front face of the SHiP spectrometer under the assumption that the dark vectors that are produced do not decay. We only accept those $V$ that are directed at the spectrometer when produced, scaled with $\varepsilon^2$ ($g_{e\mu}$) equal to one in the left (right) panel.} The novel elements of this work pertain to production within the electromagnetic cascade. Nevertheless, in order to illustrate the phenomenological relevance of these processes we have also simulated production via meson decays. We therefore organize our discussion into two subsections, first discussing ``conventional'' production, before describing the production mechanisms that operate in an electromagnetic cascade. 

Before considering production modes, let us first specify how the dark vectors couple to various SM particles. When considering a dark photon, we assume a coupling to any electrically charged particle $\psi$ proportional to $\varepsilon e Q_\psi \bar{\psi} \gamma_\mu \psi V^\mu$ with $\varepsilon$ a constant and $Q_\psi$ the electric charge of $\psi$. For the gauged lepton number, $L_\alpha-L_\beta$,  models with a coupling constant $g_{\alpha\beta}$ we take the coupling to $\ell_\alpha$ and $\ell_\beta$ to be given by $\pm g_{\alpha\beta}$, whereas the coupling to all other particles is (again) determined by kinetic mixing, with  $\varepsilon_{\alpha\beta}(q^2) \propto e g_{\alpha\beta}$.
The kinetic mixing for the $L_\alpha-L_\beta$ model, $\varepsilon_{\alpha\beta}$, depends on kinematics and is given by the sum of polarization diagrams due to $\ell_\alpha$ and $\ell_\beta$ (see Eq.~(5.1) of Ref.~\cite{Berryman:2019dme} and discussion therein). Typically, $\varepsilon_{\alpha\beta}^2 \sim 10^{-3}g_{\alpha\beta}^2$. Therefore, $L_e-L_\tau$ and $L_e-L_\mu$ couple directly to electrons with strength $g_V=g_{e\left\lbrace\tau,\mu\right\rbrace}$, whereas production from neutral meson decay will be suppressed by this loop-induced kinetic mixing. 

\subsection{Meson decays}
Proton-nucleon interactions produce a  large number of secondary hadrons. Historically, in the context of dark sector searches, a conservative strategy has been followed in which only ``primary production'' (those originating from the first interaction length) is included \cite{Batell:2009di,Essig:2010gu,deNiverville:2012ij,Izaguirre:2013uxa,Batell:2014yra,Izaguirre:2015yja,Coloma:2015pih,Alexander:2016aln,deNiverville:2016rqh,Magill:2018jla,Berlin:2018pwi,MiniBooNEDM:2018cxm,Magill:2018tbb,Berryman:2019dme,Plestid:2020kdm,Berlin:2020uwy,Batell:2020vqn,DeRomeri:2019kic,Kelly:2021xbv,Breitbach:2021gvv,Dev:2021qjj,Blinov:2021say,Gori:2022vri,Coloma:2023adi}. We follow this convention for mesons  to allow a direct comparison with existing sensitivity studies (i.e., we do not simulate the hadronic cascade and the attenuation of the primary beam in the target). 
These mesons can directly decay into dark vectors, $\mathfrak{m} \to \gamma V$.
Additionally, their SM decays $\mathfrak{m} \to \gamma \gamma$ generate a photon flux to seed an electromagnetic cascade and we include all dark sector production therein. \texttt{PETITE} takes a primary meson as an input and simulates its decay into a pair of photons, each of which initiates the EM cascade~\cite{Blinov:2024pza}.

We use \texttt{PYTHIA-8.309}  \cite{Bierlich:2022pfr} to simulate the production of light neutral mesons with QCD flag \texttt{SoftQCD:all} turned on, generating samples of $\mathfrak{m} = \pi^0$, $\eta^{(\prime)}$, and $\omega$
We decay the mesons and track their decay products or EM secondaries through the target and the detector; details of the geometry are discussed in \cref{sec:geometry}. 

In addition to $\pi^0$, $\eta^{(\prime)}$, and $\omega$, we also consider charged mesons. These particles can be important sources of dark vectors in leptophillic models. For example, in Ref.~\cite{Berryman:2019dme} it was found that $K^+ \rightarrow \ell^+ \nu_\ell V$ was the dominant production channel in the DUNE proton beam dump for leptophillic models. Importantly, unlike accelerator neutrino experiments, SHiP does not have a meson decay pipe following the beam stop and instead has a hadron absorber. As a result charged meson yields are suppressed by their interaction length divided by their decay length in the lab frame; a similar feature was recently noted in the context of  DarkQuest \cite{Blinov:2024gcw}. 
Although unimportant for the models of focus in this work, these could provide significant production in the $L_\mu - L_\tau$ gauge boson; we will explore this in detail in future work.

\subsection{Electromagnetic cascade}
We now turn to new physics particles produced in the ensuing electromagnetic cascade. High energy photons in a proton beam dump come dominantly from $\pi^0\rightarrow \gamma \gamma$ with the parent $\pi^0$ being produced by the inelastic scattering of a high energy proton from the beam. We use the same modelling of $\pi^0$ production described above (we do not include daughter photons from $\eta$, $\eta'$, and $\omega$, which are negligible). 

Once a shower is initiated by a high energy photon, the cascade proceeds via six reactions: bremsstrahlung, pair production, Compton scattering, M\o ller scattering, Bhabha scattering, and annihilation. These processes are complimented by quasi-continuous multiple Coulomb scattering, and energy losses due to ionization. All of the above reactions/processes are included in \texttt{PETITE}~\cite{Blinov:2024pza} which we use to simulate the shower's development. 

After generating an electromagnetic cascade, a Monte Carlo simulation of dark vector production can be carried out by ``dressing'' each vertex in the shower with an associated probability for the production of a dark vector~\cite{Blinov:2024pza}. The sampling of final-state kinematics is performed using the appropriate matrix element for dark vector production.

\subsection{Resonant annihilation}
As we discuss below, the gain in sensitivity (relative to published estimates considering only meson-decay production) that we project for SHiP  can be ascribed almost entirely to $e^+e^-\rightarrow V(\gamma)$. We therefore highlight this reaction in particular, emphasizing its treatment within \texttt{PETITE}~\cite{Blinov:2024pza}. The cross section for this process is resonant in nature and is highly peaked at center of mass energies close to the dark vector mass, $E_{e^+}\sim m_V^2/2m_e$.
The original implementation of $e^+e^-\rightarrow V(\gamma)$ in \texttt{PETITE} treats the electron as at rest, and includes initial state radiation via a QED parton distribution function (i.e., we have implemented a radiative return routine). Recent work has emphasized the role of momentum fluctuations due to atomic binding 
 in broadening narrow resonances \cite{Arias-Aragon:2024qji}. We have implemented a model of atomic binding in \texttt{PETITE} to investigate whether our results are sensitive to these effects. We find that the impact of atomic binding is small for SHiP, however for completeness we discuss our implementation in detail. 

In \cref{sec:resonant} we present a derivation of the tree-level cross section for a positron with momentum $\vb{k}$ annihilating  an electron in a fixed atomic orbital. The derivation follows the methods from Refs.~\cite{Plestid:2024jqm,Plestid:2024xzh}. The result differs slightly from previous expressions in the literature\footnote{Our result starts from the reaction of a positron scattering on an atom. This is different from the approach in Refs.~\cite{Celentano:2020vtu,Marsicano:2018glj,Marsicano:2018glj,Nardi:2018cxi,Marsicano:2018krp,Arias-Aragon:2024qji} where the $e^+e^-\rightarrow V$ cross section $\sigma(s)$, with $s$ a function of $\vb{p}$, is averaged over $\vb{p}$. See \cref{sec:resonant} for a more detailed discussion. \label{footnote-binding}} and is given by 
\begin{align}
    \label{sigma-def-atomic-binding}
     \sigma^{(0)}(\vb{k}) = &\frac{1}{4m_e\sqrt{\omega^2-m_e^2}} \int \frac{\dd^3p}{(2\pi)^3} 
     \frac{1}{2E_e}\frac{1}{2E_V}   \\
        &\times (2\pi)\delta(\omega+m_e -\epsilon_A -E_V) 
        |\psi_A(\vb{p})|^2 \langle |{\sf M}|^2 \rangle~. \nonumber 
\end{align}
where $\epsilon_A$ is the binding energy for the atom,  $\psi_A(\vb{p})$ the bound-state electron wavefunction, $\omega$ the energy of the incident positron, and $E_V$ the energy of the final-state vector. The matrix element ${\sf M}$ is calculated using free-electron states and is given explicitly by 
$\langle |{\sf M}|^2 \rangle =  g_V^2 [m_V^2+2m_e^2]$. 

While \cref{sigma-def-atomic-binding} has been derived for a fixed atomic orbital, its generalization to a many-electron atom is straightforward. The wavefunction squared $|\psi_A(\vb{p})|^2$ is replaced by a spectral function $S(\epsilon,\vb{p})$ \cite{Plestid:2024jqm,Plestid:2024xzh}. If all quantities are approximated by their $\epsilon\rightarrow 0$ limit, then one can use $\int \dd \epsilon S(\epsilon,\vb{p})= n(\vb{p})$, and one effectively replaces $|\psi_A(\vb{p})|^2\rightarrow n(\vb{p})$ where $n(\vb{p})$ is the momentum distribution of the atom. This distribution can be measured experimentally or computed from first principles \cite{Arias-Aragon:2024qji}.

In \texttt{PETITE} we have modeled the atomic momentum distribution with a sum of $1s$ hydrogenic orbitals $|\psi_{1s}(\vb{p})|^2\propto 1/(\vb{p}^2 + \Lambda^2)^4$ 
with $\Lambda = Z_{\rm eff} \alpha m_e$. This choice mismodels the shape of the closed-shell sum momentum distribution, $\sum_{\ell,m} |\psi_{n\ell m}(\vb{p})|^2$, however it offers a flexible and tuneable model that can be used to estimate the size of atomic binding corrections. As we discuss below we find that these corrections are relatively small, and do not qualitatively change our results. Single-photon emission is added on top of this distribution using  the leading-order $e\rightarrow e\gamma$ splitting function. When modelling molybdenum we compute the annihilation off of each atomic orbital using a separate $Z_{\rm eff}$ \cite{Slater:1930APS} but the same $1s$-orbital model mentioned above. 

The simple $1s$-orbital momentum distribution leads to an analytic expression for the resonant annihilation cross section (see \cref{sigma-res-LO-and-NLO} below); this is computationally efficient and convenient. To compute the radiative tail we use a fixed-order splitting function. The simple form of the tree-level cross section (assuming a $1s$ hydrogen-like wavefunction) allows for an analytic expression for the radiative tail as well. The result is that the $0\gamma$ and $1\gamma$ cross sections can be written as
\begin{align}
    \label{sigma-res-LO-and-NLO}
    \sigma^{(0)}(\vb{k}) &= g^2 [m_V^2 + 2m_e^2]\times \frac{2}{3 m_e\Lambda}
    \frac{1}{\vb{k}^2} \times \frac{1}{\qty((a-b)^2+1)^3}~, \nonumber \\
    \sigma^{(1)}(\vb{k}) &= g^2 [m_V^2 + 2m_e^2]\times \frac{2}{3 m_e\Lambda}
    \frac{1}{\vb{k}^2} \times \frac{\beta}{2}\mathcal{I}(a,b)~,
\end{align}
where $a=m_e/\Lambda$, $b=m_V^2/(2 k \Lambda)$, and $\beta= 2\alpha/\pi \qty(\log(s/m_e^2)-1 )$. The function $\mathcal{I}(a,b)$ is defined in \cref{sec:resonant}. The resonant nature of the cross section can be seen in the form of the functions in \cref{sigma-res-LO-and-NLO} which peak when $E_{e^+}\sim m_V^2/2m_e$.

A similar strategy can be pursued to compute atomic effects in dark Compton scattering, where a splitting function for $\gamma \rightarrow e^+ e^-$ is convolved with the resonant annihilation cross section. In this partonic level description the photon fluctuates into a high energy electron-positron pair. The positron annihilates resonantly on a bound atomic electron, while the electron is emitted in the final state. The resulting cross section for an incident photon with momentum $\vb{k}$ is given by 
\begin{equation}
    \sigma(\vb{k}) = g^2 [m_V^2 + 2m_e^2]\times \frac{2}{3 m_e\Lambda}
    \frac{1}{\vb{k}^2} \times \frac{\beta}{4}\mathcal{J}(a,b)~,    
\end{equation}
where $\mathcal{J}(a,b)$ is defined in \cref{sec:resonant}.

We find that the broadening of the resonance from the atomic momentum distribution has only minor effects on the sensitivity of SHiP to dark vectors. Notably this occurs at the largest masses, where the primary effect of the atomic wavefunction is to lower the reaction threshold ({\it cf.} discussions in Ref.~\cite{Arias-Aragon:2024qji}). Since proton beam dumps generate broad spectra of positrons, this effect is less pronounced than in e.g., the 280~{\rm MeV} PADME beam \cite{Darme:2022zfw}. For dark Compton scattering the inclusion of wavefunction broadening is even less important than for resonant annihilation. Nevertheless, we do find that including the atomic wavefunction slightly enhances sensitivity. 

\section{The Lifetime Frontier \label{sec:Lifetime}}
\begin{figure*}[t]
\centering
\includegraphics[width=\textwidth]{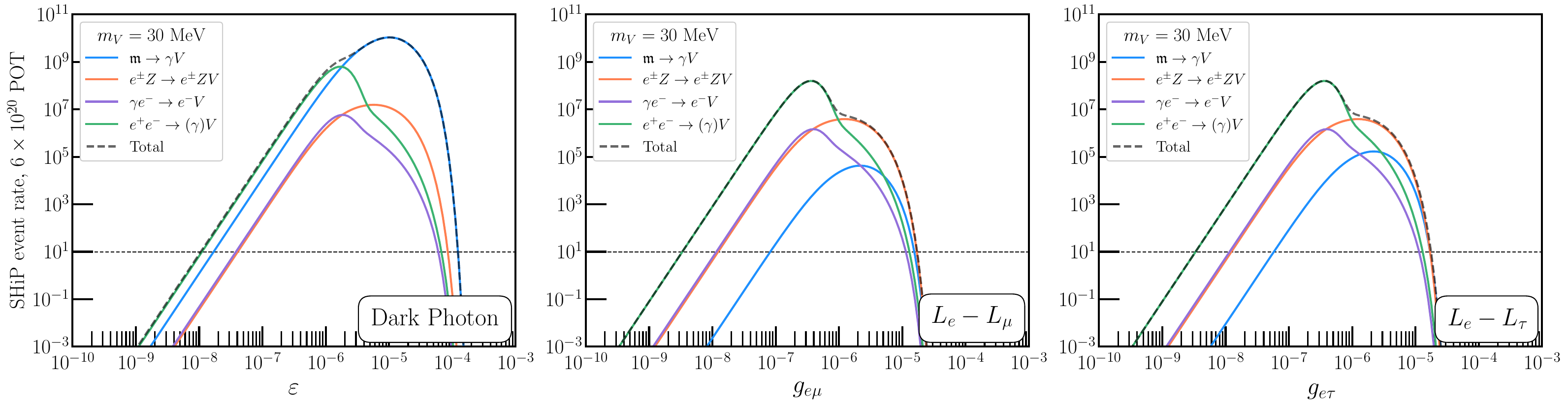}
\caption{SHiP event rate (assuming $6\times 10^{20}$ protons on target and $m_V = 30$ MeV) for three different models: dark photon and two electrophilic gauge bosons. Each panel displays the contribution to the total event rate (black, dashed) from four different processes: meson ($\pi^0$, $\eta$, $\eta^\prime$, $\omega$) decays (blue), $e^\pm$ bremsstrahlung (orange), $e^+$ annihilation (green), and Compton upscattering (purple). In each panel, as the $x$-axis coupling parameter is increased, the lifetime of the decaying particle shrinks and the relative importance of the four processes varies. We restrict our signal events to $E_V > 200$~MeV (see~\cref{subsec:EnergyCuts} for further discussion).
\label{fig:coupling-plot}}
\end{figure*} 

The expected number of dark boson decays (e.g., $V\to e^+e^-$) that can be seen at SHiP is determined by an interplay between their production cross section, $\sigma$, branching fraction to visible states and the likelihood for them to decay within the detector. The latter probability is in turn dictated by the dark vector's lifetime, $\tau$, and its typical boost in the lab frame, as well as details of the detector geometry. The boost depends on the available beam energy and the production mechanism. In general, ignoring detector geometry for now, the expected number of dark vectors decaying between the front of a decay pipe, distance $L$ from the production point, and the back of the detector, a further distance $\ell$ away is
\begin{equation}\label{eq:Nobsscaling}
    N_{\rm obs.}\propto \sigma \left(e^{-L/\gamma\beta c\tau_V} - e^{-(L+\ell)/\gamma\beta c\tau_V}\right)~.
\end{equation}
At large couplings the production rate is high but the lifetime is so short that the number of dark vectors that survive to the detector becomes exponentially small, unless they are very boosted in the lab frame. 
In the opposite limit, as the dark vector's couplings become small ($\varepsilon$ or $g_{\alpha\beta}\rightarrow 0$), the production rate decreases and the dark vector's lifetime becomes very long. At sufficiently small coupling, the lab-frame decay length $\lambda_V = \gamma\beta c\tau_V\propto g_{\alpha\beta}^{-2}$ becomes longer than SHiP's decay pipe, $\lambda_V\gg 50~{\rm m}$, and the probability that a dark vector produced in the beam dump decays within the decay pipe scales as $\ell/\lambda_V$. The time-dilation associated with the boost causes the lab-frame decay length of dark vectors to scale as $\lambda_V \propto E_V/m_V$ such that, at small coupling, lower-energy particles are more likely to decay inside the decay pipe. The interplay between production, decay, and boost can be seen by comparing the reach for different classes of dark vector models.

In models with electrophilic couplings, such as $L_e-L_\mu$ (shown in the right column of~\cref{fig:energy-spectrum}) and $L_e-L_\tau$, the electromagnetic cascade dominates over meson production for all vector energies. 
For more democratic models such as the dark photon (left column of \cref{fig:energy-spectrum}), the dominant process depends on the proper vector decay length: at long decay lengths, the lab frame probability for $V$ to decay in the fiducial region is enhanced for smaller boosts (smaller $V$ energies). We therefore expect processes that produce $V$'s with lower typical energies to be more relevant at small lifetimes.
Examining the left column of \cref{fig:energy-spectrum}, we see that electromagnetic production results in substantially lower energies. Furthermore, at low dark vector mass, the integrated yield for resonant annihilation $e^+e^-\rightarrow V(\gamma)$ is comparable to that stemming from $\pi^0\rightarrow \gamma V$. As a result, which of these two components of the flux dominates the number of events at the SHiP detector downstream depends on $\lambda_V$ and therefore $\varepsilon$.

While \cref{fig:energy-spectrum} shows the expected flux of dark vectors entering the front face of the SHiP spectrometer it does not take into account the probability of decay within the detector. This effect is shown in \cref{fig:coupling-plot} where we plot the number of $V$ decays in the decay volume as a function of the relevant coupling for a $30~{\rm MeV}$ dark vector. As we will discuss in~\cref{subsec:EnergyCuts}, the minimum energy of vectors allowed in the analysis here is $200$~MeV. The effects discussed above can be seen in these plots. For instance, in the dark photon panel (left), we can see the cross-over between large coupling where the dark photon is short lived and $\pi^0\rightarrow \gamma V$ dominates to low coupling where it is longer lived and $e^+e^-\rightarrow V(\gamma)$ dominates. At very small coupling the slowest-moving $V$s dominate the rate and the number of events at SHiP, from all processes, scales as the fourth power of coupling. In hadrophobic models, bremsstrahlung production is enhanced relative to meson decay, so $e^\pm Z\rightarrow e^\pm Z V$ dominates at intermediate couplings as illustrated in the middle and right panels of \cref{fig:coupling-plot}. 

In summary, we find that electromagnetic cascades perform two roles when acting as a new physics factory. First, they have a large multiplicity of $e^-$, $e^+$, and $\gamma$, all of which can produce a flux of dark vectors downstream. Second, the resultant flux is substantially {\it less boosted} than the flux produced by primary interactions. For large lifetimes, these less-boosted particles are more likely to decay inside the volume of interest, and therefore to result in a signal. Both of these effects lead to an enhancement in the sensitivity of beam dump facilities to visibly decaying long-lived particles once electromagnetic production is included.

\begin{figure*}[t]
\begin{center}
\includegraphics[width=\textwidth]{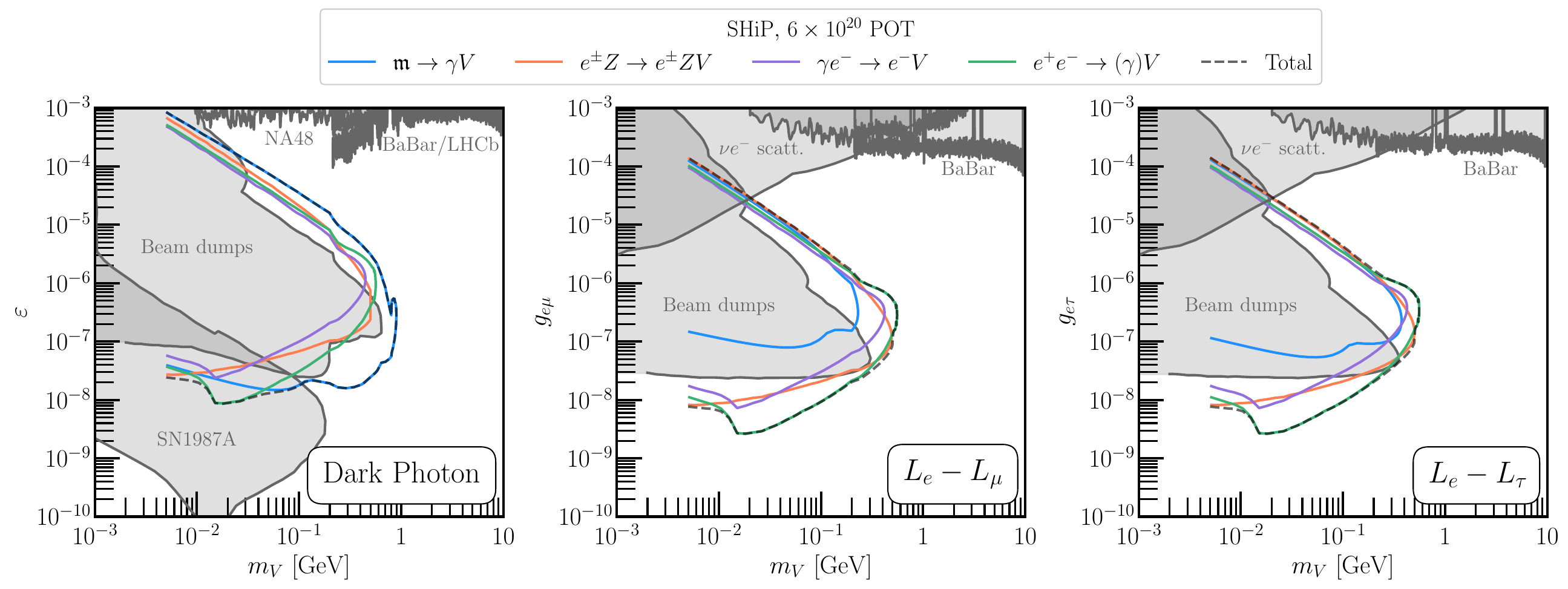}
\end{center}
\caption{Projected ``meson only'' (i.e., neglecting proton-bremsstrahlung and Drell-Yan production) SHiP sensitivity contours for secluded dark photon and two leptophilic gauge bosons at $N_{\rm evt.} \geq 10$ and $6\times 10^{20}$~protons on target. All injected photons come from the same meson decays used for $\pi^0\rightarrow \gamma V$. For each panel, the different-colored contours correspond to different production mechanisms and the gray dashed line indicates the overall SHiP sensitivity. Shaded regions correspond to existing constraints from different sources -- see labels and text for more detail.\label{fig:SHiP_projections}}
\end{figure*}

\section{SHiP Sensitivity \label{sec:Sensitivity}} 

We now use our model of vector production to study the projected sensitivities for the SHiP experiment. As a benchmark, we assume $6 \times 10^{20}$ protons on target (POT) \cite{Albanese:2878604} and define our sensitivity contours by  $N_{\rm sig.} \geq 10$. We find qualitatively similar results as Refs.~\cite{Bauer:2018onh,SHiP:2020vbd} upon correcting for the different assumptions of POT, detector geometry, and exclusion thresholds when calculating sensitivity contours (2.3 events in \cite{SHiP:2020vbd} vs.~10 events in \cite{Bauer:2018onh}). We do, however, find some quantitative differences (mass reach and coupling sensitivity changes by factors of a few) which we will explore in detail in future work. As expected, electromagnetic secondary production dominates over a wide range of parameter space for electrophillic models, and competes at lower masses and lower couplings for the dark photon model. 

Many searches for BSM physics at SHiP are expected to be close to background free \cite{Aberle:2839677}.  We do not expect the search for low energy dark vectors (from the EM cascade) to be any different, and preliminary analyses show that there is not an excessive pile up of backgrounds with low track momenta \cite{MachadoSantosSoares:2765979}.  However, the exact background rate is subject to change.  As a measure of the impact of background on the sensitivity of SHiP we present in \cref{app:varyingNsig} the effect of requiring more signal events.  Going from $N_{\rm sig.}=10$ to $N_{\rm sig.}=1000$ reduces the bound on $g$ or $\varepsilon$ by approximately a factor of 3.

\subsection{Updated geometry \label{sec:geometry} }
Past studies~\cite{Bauer:2018onh,SHiP:2020vbd} have been performed using the SHiP designs from the original technical proposal \cite{SHiP:2015vad}. This specification included a downstream spectrometer at $z=120$ m, and a decay pipe from $z = 60$ m to $z = 110$ m (with the target centered on $z=0$). 
The version of the experiment that was approved in March 2024 
is more compact~\cite{Aberle:2839677,Albanese:2878604}, requiring a re-analysis of many existing projections. For long-lived particle signals considered here, the relevant detector subsystem is the hidden sector decay spectrometer (HSDS). Its new geometry has a decay region between $z=33.5$ m and $z = 83.5$ m, followed by a charged particle tracker, timing detector, electromagnetic and hadronic calorimeters. The tracker has an aperture of $4$ m by $6$ m determining the angular acceptance of potential decays. These requirements are adopted in generating the expected event rates demonstrated in~\cref{fig:coupling-plot}, as well as in the following results. More details on the determination of the impact of detector geometry are presented in \cref{app:Efficiency}. 

\subsection{Energy cuts at SHiP}\label{subsec:EnergyCuts}


In addition to an acceptance penalty related to detector geometry and the requirement that the long lived vectors decay visible final states within the SHiP detector, discussed above, there are further suppressions of the signal rate related to detector efficiencies. While a full detector simulation is beyond the scope of this work, here we investigate how the existence of an energy threshold to observe the final state electrons impacts the reach at SHiP. There are two main systems involved in the identification of electrons and the reconstruction of their four momentum: the straw tracker and the electromagnetic calorimeter. There are proposals for future upgrades to the SHiP detector such as a downstream Liquid Argon Time Projection Chamber (LArTPC) \cite{Albanese:2878604} which could substantially improve the efficiency of reconstructing low energy electrons.

The electron-positron pair from a dark vector decay in the decay volume first pass through the spectrometer, which is inside a large dipole magnetic field. The magnetic field has an approximate Gaussian profile in the beam direction, with peak field of $0.14$ T and an average over the $10$ m length of the spectrometer of $0.08$ T. 
In order for an electron's gyroradius to be larger than the radius of the spectrometer, and thus leave hits in all the tracking stations, its energy must be above $115$ MeV. This is the requirement for an electron passing down the center of the spectrometer, a lower energy electron whose initial motion is opposite to its bending may also make it through the tracker. With dedicated algorithms it may also be possible to reconstruct spiralling tracks. With a sufficient number of hits, the straw tracker is expected to have excellent momentum resolution, at these energies the error is about $0.5\%$ \cite{SHiP:2015vad}. 

The electromagnetic calorimeter is nominally proposed to measure the energy of $\gamma/e$ over a broad range of energies from $300$ MeV to $70$ GeV. 
The energy resolution  is expected to be $\sigma(E)/E\sim 0.06 \sqrt{\GeV/E}$ and is about $10\%$ at $300$ MeV. The resolution of the analog-to-digital converter (ADC) is about $20\ \MeV$ and below about $125\ \MeV$ this becomes the limiting factor. It is expected that during spills there will be $10^5~ \mu/\mathrm{sec}$, with average momentum of $10\ \GeV$, hitting the calorimeter. Each of these muons deposits the same energy as a $400\ \MeV$ $\gamma/e$. This background presents a challenge to looking for smaller-energy electromagnetic deposits, but it may be possible to see below this noise level by looking for coincidences in the muon system, which is downstream of the ECal, as well as looking for the different tracks in the spectrometer, or by using timing.

In the lab frame, the electron and positron do not share the energy of the dark vector equally and the energy threshold requirement must be applied to both final state particles. Given the uncertainties on the exact value of this threshold we instead choose to require that the energy of the dark vector is above twice the expected electron/positron threshold. For our main result, we take this threshold for the $e^\pm$ to be $E^{e^\pm}_{\mathrm{thresh.}}=100~\MeV$ and thus require $E_V\ge E^V_{\mathrm{thresh.}}= 2\times(100~\MeV)$. In \cref{app:energythreshold} we present results for the expected reach under various alternative assumptions for this threshold, in particular $E^V_{\mathrm{thresh.}}=0\MeV,~100\MeV,~300\MeV,~1\GeV$. We view $E^V_{\mathrm{thresh.}}=200\MeV$ as a relatively conservative requirement since there are many approaches (timing, decreasing the B-field, LArTPC, etc) which may allow reconstruction of lower energy dark vectors, perhaps even down to $50~\MeV$ \cite{Nico:chat}, corresponding to electrons around $25~\MeV$.

\subsection{Results and discussion}

To compare with other experimental results and projections for the three models of interest, we divide the sensitivities that we produce into a ``mesons-only'' projection, as well as a ``full'' projection that includes all production modes discussed in this work. We explicitly do not include proton-bremsstrahlung and Drell-Yan/direct perturbative QCD production estimates in our projection. These processes are not important for the lower-mass sensitivity of SHiP, however they dominate sensitivity in the dark photon model for masses $m_V \gtrsim$ 500~MeV, see, e.g., Ref.~\cite{SHiP:2020vbd} for further discussion. 

\Cref{fig:SHiP_projections} summarizes our main result, where we display projections at SHiP (10-event contours), comparing a secluded dark photon and two leptophilic ones with different production modes. For a secluded dark photon, positron annihilation pushes sensitivity at lower couplings down to $10^{-8}$. Although some of this parameter space is already excluded by supernova cooling bounds \cite{Chang:2016ntp}, our results suggest that SHiP will provide the strongest terrestrial constraints on dark photons in the ${\sim}10~{\rm MeV}$ mass range surpassing SLAC-E137 constraints \cite{Bjorken:1988as}. In the dark photon parameter space, we also compare sensitivity against existing beam-dump facilities~\cite{Bauer:2018onh}, as well as searches from NA48~\cite{NA482:2015wmo}, BaBar~\cite{BaBar:2014zli,BaBar:2016sci}, and LHCb~\cite{LHCb:2019vmc}.

For electrophilic models $L_e-L_\mu$ and $L_e-L_\tau$, we compare again against the existing constraints from beam-dump experiments~\cite{Bauer:2018onh} and BaBar~\cite{BaBar:2014zli,BaBar:2016sci} (utilizing \texttt{darkcast}~\cite{Ilten:2018crw,Baruch:2022esd} in doing so). We also display constraints from neutrino-electron scattering derived for TEXONO~\cite{TEXONO:2009knm,Bilmis:2015lja, Lindner:2018kjo, Bauer:2018onh} and Super-Kamiokande~\cite{Wise:2018rnb}. For SHiP, vector production via meson decay is suppressed by loop-induced couplings, allowing the electromagnetic secondaries to dominate particularly in the low coupling regime. As a result, sensitivities are enhanced relative to the meson-decay contributions -- more than an order of magnitude in coupling and three times in mass. We would like to emphasize that upon including electromagnetic production, a proton fixed-target experiment like SHiP is able to probe a large portion of parameter space that was not previously explored by current \textit{electron} beam dump experiments~\cite{Bauer:2018onh}. The loss of sensitivity for $m_V\ltap 10\,\mathrm{MeV}$ corresponds to resonant production falling below the $E_V > 200$~MeV energy threshold mentioned above. Better sensitivity at lower masses can be attained with a lower energy cut (see~\cref{app:energythreshold}); this will require dedicated study by the collaboration. 

Sensitivity in the $L_e - L_\mu$ and $L_e - L_\tau$ models is very similar -- differences arise in comparing the sensitivity reach from meson decays (driven by the loop-induced kinetic mixing, which is slightly different between the two models) and in the available final-states of decay. For $m_V \geq 2m_\mu$, an $L_e - L_\mu$ gauge boson may decay into di-muon pairs whereas an $L_e - L_\tau$ cannot. This also impacts the lifetime as a function of the boson's mass and gauge coupling.

\subsection{Electrophobic models \label{sec:Lmu-Ltau}}
As mentioned in the introduction, the electromagnetic cascade is less impactful as a resource for electrophobic models. An interesting example is the aforementioned gauged $L_\mu-L_\tau$ model. This example is both electrophobic and hadrophobic with a sizeable branching ratio into neutrinos. As a result the $L_\mu-L_\tau$ gauge boson is notoriously difficult to search for. 

We have conducted a preliminary investigation into the sensitivity of the SHiP experiment to $L_\mu-L_\tau$ models and find that sensitivity is {\it significantly} weaker than previously estimated in the literature. The main differences between our analysis and existing studies arise from the following features:
\begin{itemize}
    \item Sensitivity cannot be simply recasted from dark photon models because the ``floor'' and ``ceiling'' of the sensitivity projections are too close together. 
    \item Electromagnetic secondaries can provide improvements in sensitivity as compared to meson decays, but only if the reach of the experiment is sufficiently strong such that the long-lived limit is being probed. For SHiP we project a weaker reach where this does not occur.
    \item Production modes beyond meson decays and the electromagnetic cascade are important, and should be re-evaluated. This includes previously neglected production modes including $D$ meson decays and muon bremsstrahlung in the hadron absorber. 
\end{itemize}
The qualitative features of $L_\mu-L_\tau$ are sufficiently different from dark photons, $L_e-L_\tau$, and $L_e-L_\mu$ that we defer a detailed discussion to future work. Since the sensitivity is largely dictated by processes other than the electromagnetic cascade the discussion of $L_\mu-L_\tau$ is largely independent of the discussion presented above. 

\section{Discussion \& Conclusions \label{sec:Conclusions}}
We have studied the production of dark vectors at the SHiP experiment. We find that electromagnetic cascades offer an underappreciated resource in high energy proton beam dumps; this mirrors similar conclusions from Refs.~\cite{Celentano:2020vtu,Dutta:2020vop,Brdar:2020dpr,Capozzi:2021nmp,Brdar:2022vum}. A novel aspect of our findings is that the secondary electromagnetic cascade is important even when the total yield of dark photons is dominated by meson decays. This can occur when lifetimes are very long, and the softer spectrum from cascade processes such as $e^+e^-\rightarrow V(\gamma)$ is advantageous due to the smaller time dilation experienced by the unstable dark vectors. 


Three major consequences of our analysis are:
\begin{itemize}
    \item \textbf{Enhanced sensitivity at small couplings:} The electromagnetic cascade can increase signal rates in the SHiP downstream detector by orders of magnitude. This translates to an improvement in reach by factor $\sim 5-10$ in coupling $\varepsilon$ or $g_{\alpha\beta}$.
    \item \textbf{Extended mass reach:} For electrophilic models we find that the mass reach of SHiP with resonant annihilation $e^+e^-\rightarrow V(\gamma)$ included is roughly $3\times$ larger than ``meson-only'' production. It would be interesting to compare projections including proton bremsstrahlung and direct QCD production. 
    \item \textbf{New search strategies:} Many of the gains in sensitivity we have found stem from portions of the $e^+e^-$ flux with low energies. We have estimated (what we expect to be) realistic energy cuts at SHiP, however our study motivates efforts to better understand the performance of the detector and background rejection systems in the ${\sim}100~{\rm MeV}$ energy range. 
\end{itemize}

Including electromagnetic cascades substantially enhances the sensitivity of experiments like SHiP to feebly interacting particles coupled to electrons or photons with masses below $\sim 50-300 ~{\rm MeV}$ (depending on the model considered). Whether or not electromagnetic cascades dominate projected sensitivities is a model dependent question that depends crucially on the lifetime of the outgoing particles. Nevertheless, inclusion of the cascade is a resource which can only enhance the sensitivity of SHiP and other beam dump experiments searching for feebly interacting particles, and should be included in future projections for high energy beam dump experiments.

\section*{Acknowledgements} 
We thank P.A.N. Machado for collaboration in early stages of this work, and E.~Nardi for detailed correspondences relating to atomic binding effects. We are also grateful to Stefania Gori for useful discussions, and Nicola Serra regarding energy thresholds at SHiP and the experiment's updated nominally planned number of protons on target. 
Part of this research was performed at the Aspen Center for Physics, which is supported by National Science Foundation grant PHY-1607611. RP is supported by the Neutrino Theory Network under Award Number DEAC02-07CH11359, the U.S. Department of Energy, Office of Science, Office of High Energy Physics under Award Number DE-SC0011632, and by the Walter Burke Institute for Theoretical Physics. NB acknowledges the support of the Natural Sciences and Engineering Research Council of Canada (NSERC). This research was enabled in part by support provided by the BC DRI Group, Compute Ontario and the Digital Research Alliance of Canada (\href{http://alliancecan.ca}{alliancecan.ca}).

\appendix

\section{Atomic binding corrections \label{sec:resonant} }
Motivated by previous work \cite{Celentano:2020vtu,Marsicano:2018glj,Marsicano:2018glj,Nardi:2018cxi,Marsicano:2018krp,Arias-Aragon:2024qji}, we have implemented a model of atomic binding corrections for resonant annihilation $e^+ e^- \rightarrow V(\gamma)$. Our formalism differs somewhat from existing treatments in the literature, and so in this Appendix we present a self-contained derivation of the formulae used in the updated version of \texttt{PETITE}. 

\begin{figure}[b]
    \centering
    \includegraphics[width=\linewidth]{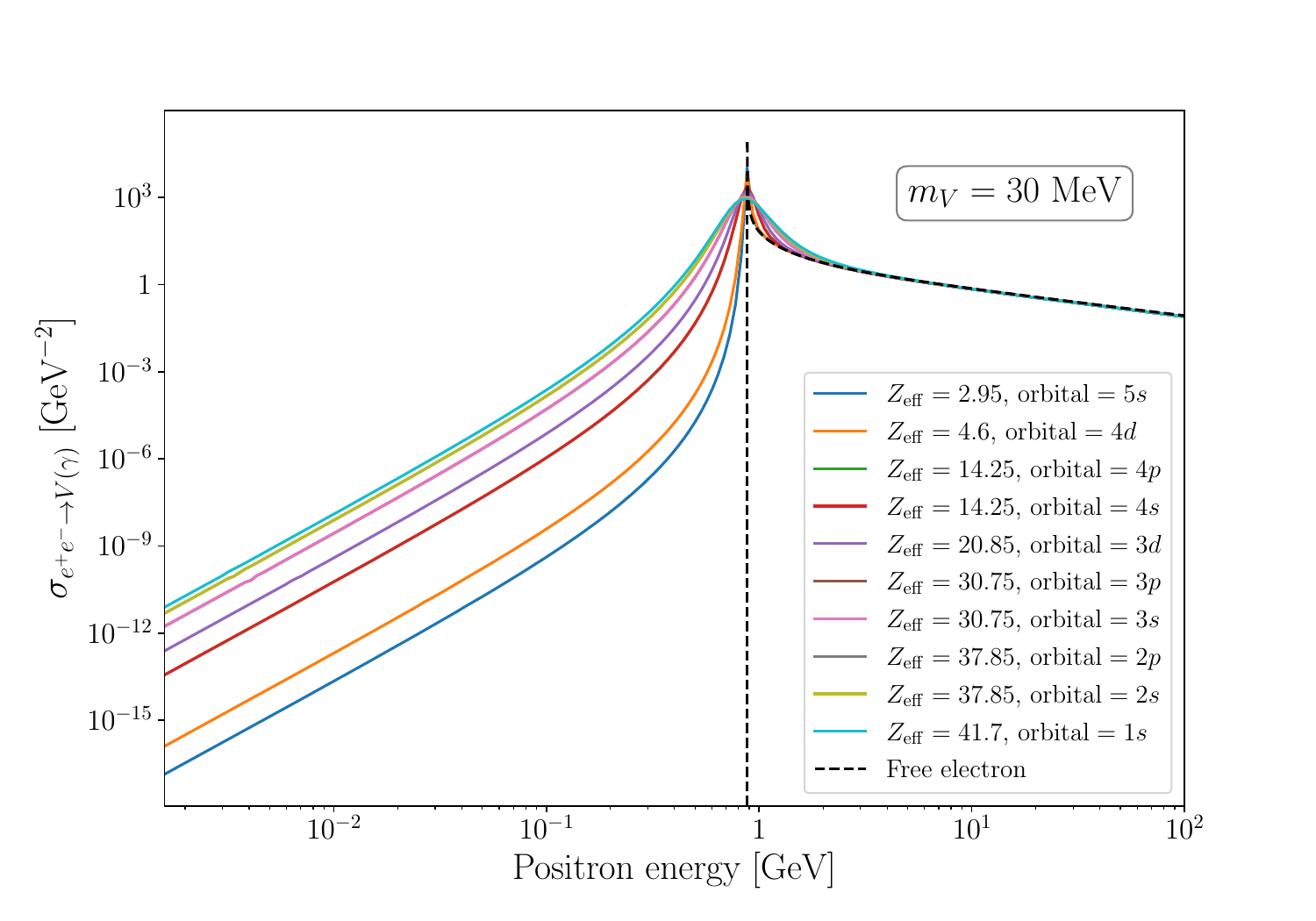}
    \caption{Effect of effective nuclear charge $Z_\mathrm{eff}$ (and the corresponding electron orbital specification) on dark annihilation cross section, with vector mass fixed to 30 MeV. The $Z_\mathrm{eff}$ is calculated using Slater's rule~\cite{Slater:1930APS} while the shape of the momentum distribution for each orbital is given by \cref{1s-orbital}.
    \label{fig:xsec_zeff}}
\end{figure}

We will model the atom as a single electron bound to a heavy nucleus, $B$, of charge $Z$ (in practice $Z\equiv Z_{\rm eff}$ which can vary for different atomic orbitals). The impact of different choices of $Z_{\rm eff}$ on the cross section is shown in \cref{fig:xsec_zeff}. We will consider the atom to have definite mass $m_A$, and the heavy nucleus to have definite mass $m_B$.

\subsection{Tree-level cross section \label{sec:tree-level-atomic} }
We begin by considering the scattering off the atom as a whole. The treatment here closely mirrors Ref.~\cite{Plestid:2024xzh}. In the absence of initial state radiation, the reaction of interest may be written in the lab frame as 
\begin{equation}
    e^+(\vb{k}) ~+~ A(\vb{0}) \rightarrow  V(\vb{q}) + B(\vb{p}_B)~. 
\end{equation}
This is a $2\rightarrow 2$ scattering process. The outgoing momentum of the nucleus $B$ is set by the atomic wavefunction $|\vb{p}_B| \sim Z\alpha m_e$. We will neglect the kinetic energy of $B$ (it is very small $T_B\sim Z^2\alpha^2 m_e^2/m_B$) in what follows by taking $E_B\simeq m_B$. The mass difference between the initial atom and the outgoing nucleus defines the binding energy, $\epsilon>0$, 
\begin{equation}
    m_A-m_B = m_e -\epsilon~. 
\end{equation}
It will be important in what follows that the initial electron has definite energy, but indefinite momentum.

Let us define the matrix element for scattering from the atom, $\mathcal{M}$, via 
\begin{equation}
    (2\pi)^4 \delta^{(4)}(\Sigma P_{AB})\mathcal{M} = \mel{V B}{\hat{\mathcal{H}}_{\rm int}}{A e^+}~. 
\end{equation}
where $\hat{\mathcal{H}}_{\rm int} = g_V V_\mu \bar{\psi}_e \gamma^\mu \psi_e$ and $\Sigma P_{AB}=[p_A + k] - [q+p_B] $ with $p_A = (m_A, \vb*{0})$ and 
$k = (\omega, \vb{k})$ in the lab frame. Then the  leading order cross section (i.e., $\sigma^{(0)}$ for $0\gamma$) for scattering from the atom may be written as 
\begin{equation}
    \label{atom-phase-space}
    \sigma^{(0)}(k) = \frac{1}{4 \mathcal{F}_A} \int \dd \Pi_V \dd \Pi_B  (2\pi)^4\delta^{(4)}(\Sigma P_{AB}) 
        \langle |\mathcal{M}|^2 \rangle~, 
\end{equation}
where $\dd \Pi_a = \dd^3 p_a / [(2\pi)^3 (2E_a)]$ is the Lorentz invariant phase-space measure and $\mathcal{F}_A=\sqrt{(p_A\cdot k)^2-m_e^2 m_A^2}$. Carrying out the integration over $\dd^3 p_V$ we find
\begin{equation}
     \sigma^{(0)}(k) = \frac{1}{4 \mathcal{F}_A} \int \dd \Pi_B\frac{1}{2E_V} (2\pi)\delta(\Sigma E_{AB}) 
        \langle |\mathcal{M}|^2 \rangle~. 
\end{equation}
To make further progress we must understand how $\langle |\mathcal{M}|^2 \rangle$ depends on $\vb{p}_B$. 

We can relate $\mathcal{M}$ to a free-electron matrix element, which we denote by ${\sf M}$, by using our knowledge of hydrogenic wavefunctions. The state $\ket{A}$ can be written in terms of a free nucleus $\ket{B}_0$ and a free electron $\ket{e}_0$ valence states as (we neglect higher Fock states that appear in the relativistic theory)
\begin{equation}
    \label{hydrogen-wf}
    \ket{A} \simeq \int \frac{\dd^3p}{(2\pi)^3}  \sqrt{\frac{2 m_A}{2E_e(p) 2 m_B}}\psi_{A}(\vb{p}) \ket{e(\vb{p})}_0 \ket{B(-\vb{p})}_0~. 
\end{equation}
Being infinitely heavy, the nucleus, $B$, effectively sources a potential for both the bound electron and incident positron. Therefore, the state $\ket{A e^+}$ contains both a bound electron wavefunction [as in \cref{hydrogen-wf}] and a continuum wavefunction for the positron. The consistent use of wavefunctions in the presence of a Coulomb field guarantees current conservation. This allows us to make the replacement 
\begin{equation}
    \label{pol-sum}
    \sum_{\rm spins} \epsilon_\mu^*(\vb{q}) \epsilon_\nu^*(\vb{q}) \rightarrow - g_{\mu\nu}  ~, 
\end{equation}
when summing over polarizations of the dark vector. We are then left to compute the contraction with $(-g_{\mu\nu})$. 

Although the positron's wavefunction is distorted (as is necessary to ensure current conservation), a simplification emerges in the high-energy limit. At high energies, the positron can be treated using the eikonal approximation, and in this limit one finds that \cite{Plestid:2024jqm}
\begin{equation}
    \int \dd\Pi_B  \langle |\mathcal{M}|^2 \rangle  =  \int \dd\Pi_B  \langle |\mathcal{M}|^2 \rangle_{{\rm free}-e^+}~.
\end{equation}
This identity only holds after integrating over the final state phase space and using the replacement \cref{pol-sum}, and relies on the heavy-$B$ limit (where the kinetic energy is negligible). 

Treating the positron as a free particle, and using $\braket{B(\vb{p}_B)}{B(-\vb{p})} = 2 m_B (2\pi)^3 \delta^{(3)}(\vb{p}_B+\vb{p})$, 
\begin{equation}
    \langle |\mathcal{M}|^2 \rangle_{{\rm free}-e^+} =  \frac{2 m_A 2m_B }{2 E_e(p_B)} |\psi_A(-\vb{p}_B)|^2 \langle |{\sf M}|^2 \rangle~.
\end{equation}
This leads to (using $\vb{p}=-\vb{p}_B$)
\begin{align}
     \sigma^{(0)}(k) = \frac{2m_A}{4 \mathcal{F}_A} \int &\frac{\dd^3p}{(2\pi)^3} 
     \frac{1}{2E_e}\frac{1}{2E_V}   \\
        &\times (2\pi)\delta(\Sigma E_{AB}) 
        |\psi_A(\vb{p})|^2 \langle |{\sf M}|^2 \rangle~. \nonumber 
\end{align}
where we have introduced the free-electron matrix element 
\begin{equation}
    \langle |{\sf M}|^2 \rangle = g_V^2\qty(2 k\cdot p+ 4m_e^2)~ ,
\end{equation}
where $k\cdot p = \omega \sqrt{\vb{p}^2+m_e^2} - \vb{k} \cdot \vb{p}$. Next we integrate over $\dd \cos\theta$ (the angle of the bound electron's momentum relative to the beam axis) using the energy conserving delta function, and the azimuthal angle of the electron, 
\begin{align}
     \sigma^{(0)}(k) = \frac{2m_A}{4 \mathcal{F}_A} \int_0^\infty &\frac{\dd |\vb{p}|  }{(2\pi)} 
     \frac{1}{2E_e}\frac{\vb{p}^2}{2 |\vb{p}| |\vb{k}|}   \\
        &\times  \Theta(|\cos\theta|-1)
        |\psi_A(\vb{p})|^2 \langle |{\sf M}|^2 \rangle~, \nonumber 
\end{align}
with the angle $\cos\theta$ fixed to equal 
\begin{align}
    \cos\theta &= \frac{m_e^2 + (m_e-\epsilon)^2+2 \omega(m_e-\epsilon)-m_V^2-\vb{p}^2}{2 |\vb{k}| |\vb{p}|} ~\nonumber \\
    &\simeq \frac{2m_e^2 +  2 \omega m_e-m_V^2 }{2 |\vb{k}| |\vb{p}|}~. 
\end{align}
The Heaviside function imposes constraints on the momentum integration 
\begin{equation}
    p_{\rm min,max} = |\vb{q}|\mp|\vb{k}|~.
\end{equation} 
These expressions disagree with those found in Refs.~\cite{Celentano:2020vtu,Marsicano:2018glj,Marsicano:2018glj,Nardi:2018cxi,Marsicano:2018krp} due to a different energy-conservation condition.\footnote{In particular, we set the dark vector's energy to be $m_e + \omega - \epsilon \simeq m_e +\omega$ as opposed to $\sqrt{m_e^2+\vb{p}^2} +\omega$ as was done in Ref.~\cite{Celentano:2020vtu,Marsicano:2018glj,Marsicano:2018glj,Nardi:2018cxi,Marsicano:2018krp}. } The treatment of kinematics agree with the updated version of Ref.~\cite{Arias-Aragon:2024qji}. There are still small differences between our expressions and those in Ref.~\cite{Arias-Aragon:2024qji} that stem from the modeling of the cross-sections on the atom as a wavefunction-averaged cross section on individual free electrons. This leads to the factor of $1/|v_A-v_B|$ in Eq.~(1) of \cite{Arias-Aragon:2024qji}, as opposed to our derivation which begins from \cref{atom-phase-space} and relates the matrix element with the atom to a matrix element on a free-electron. 

In \texttt{PETITE} we have chosen a very simple model for $\psi(\vb{p})$ that is designed to capture the typical momentum of each orbital (the most important feature for bound-state effects). We use a non-relativistic $1s$ atomic wavefunction for all orbitals (independent of principle and angular quantum numbers) but we treat the inverse Bohr radius as a tuneable parameter $\Lambda = Z_{\rm eff} \alpha m_e$. Therefore, in \texttt{PETITE} one has 
\begin{equation}\label{1s-orbital}
    |\psi_A(\vb{p})|^2  = \frac{64\pi \Lambda^5}{(p^2+\Lambda^2)^4}~. 
\end{equation}
When computing the tree-level cross section we expand in $\vb{p}/m_e$ under the integral. Up to corrections of order $\vb{p}^2/m_e^2$ we have, 
\begin{align}
    \label{tree-level-atomic}
     \sigma^{(0)}(k) = &\frac{g_V^2(m_V^2 + 2m_e^2)}{4 \mathcal{F}_e} \int_{p_{\rm min}}^\infty \frac{\dd |\vb{p}|  }{(2\pi)}  \frac{\vb{p}^2}{2 |\vb{p}| |\vb{k}|} |\psi_A(\vb{p})|^2   \nonumber \\
     =& \qty( \frac{g^2 [m_V^2 + 2m_e^2]}{8 |\vb{k}|^2 m_e} )\times \frac{16\Lambda^5 }{3(p_{\rm min}^2+\Lambda^2)^3}~
\end{align}
where $\mathcal{F}_e = m_e |\vb{k}|$, we set $\epsilon=0$ everywhere, and take the $p_{\rm max} \rightarrow \infty$ limit. One can check that in the $\Lambda\rightarrow 0$ limit this expression reduces the the free-electron cross section as it must, 
\begin{equation}
     \lim_{\Lambda \rightarrow 0} \sigma^{(0)}(k)  = \frac{g^2(m_V^2+2 m_e^2)}{4 \mathcal{F}_e}  (2\pi) \delta(s-m_V^2)~, 
\end{equation}
where $s=2m_e\omega +m_e^2$ and we have used 
\begin{equation}
   (2\pi) \delta(p_{\rm min}) = (2\pi) \frac{m_V^2}{2 m_e^2}\sqrt{1-\frac{4 m_e^2}{m_V^2}} \delta(\omega-\omega_{\rm res})~.
\end{equation}
\subsection{Radiative tail \label{sec:rad-tail} }
In the original version of \texttt{PETITE} resonant annihilation $e^+e^-\rightarrow V(\gamma)$ was treated using a Kuraev-Fadin distribution function \cite{Kuraev:1985hb,Nicrosini:1986sm} (i.e., a QED parton distribution function) for both the electron and positron. Our analysis was performed in the center-of-mass frame and boosted back to the lab frame. When considering a bound atomic electron the heavy nucleus identifies the lab-frame as ``preferred''. It is therefore helpful to understand how the radiative return analysis proceeds in the lab-frame. 

When we include an atomic wavefunction, the leading-order cross section becomes a smooth function instead of a singular (Dirac-delta) distribution. This makes the cross-section IR-safe, and permits the use of a fixed order splitting function. One can check, carefully treating the Kuraev-Fadin distributions, that they have a perturbative expansion in distributions given by
\begin{equation}
    f_e(x,s) = \delta(1-x) + \frac{\beta}{4} \qty[\frac{1+x^2}{[1-x]_+} + \frac32 \delta(1-x)] + O(\beta^2)~, 
\end{equation}
where $\beta= 2\alpha/\pi \qty(\log(s/m_e^2)-1 )$. One sees that to $O(\beta)$ the Kuraev-Fadin function is replicated by a standard fixed-order splitting function (to leading-log accuracy). In the center-of-mass frame, an $O(\beta)$ treatment of radiative return involves a splitting off either electron leg. Let us write $\sigma(k)= \sigma^{(0)}(k) + \sigma^{(1)}(k) + \ldots$. Using a fixed-order splitting function one finds, 
\begin{equation}
    \sigma^{(1)}(s) = \frac{\beta}{2} \int \dd x  \frac{1+x^2}{1-x} \qty[\sigma^{(0)}(x s) - \sigma^{(0)}(s)]~.
\end{equation}
The same parametric result is obtained in the lab-frame, but with a different parton-level interpretation. In the lab-frame only the positron can emit hard-collinear radiation. However, unlike in the center-of-mass frame, where both the electron and positron have energy $\omega_{\rm CM} \sim \sqrt{s}$, the positron in the lab frame has energy $\omega_{\rm lab} \sim s/m_e \sim \omega_{\rm CM}^2/m_e$. The higher energy in the lab-frame compensates for the absence of a splitting function for the (soft) electron. 

When calculating the radiative tail in \texttt{PETITE} we therefore make use of a single fixed-order splitting function. for consistency with the free-electron implementation we define the coefficient of the splitting function in terms of $\beta=2\alpha/\pi \qty(\log(s/m_e^2)-1 )$. For the purposes of calculating the radiative tail, it is convenient to re-write the leading-order cross section as 
\begin{equation}
    \sigma(k) \simeq g^2 [m_V^2 + 2m_e^2]\times \frac{2}{3 m_e\Lambda}
    \frac{1}{\vb{k}^2}\frac{1}{\qty((a-b)^2+1)^3}~,
\end{equation}
where
\begin{align}
    a&=m_e/\Lambda~,\\
    b&=m_V^2/(2 k \Lambda)~.
    \label{eq:ab_coef}
\end{align}
Then we find that ({\it cf.} \cref{sigma-res-LO-and-NLO} of the main text), 
\begin{align}
    \label{analytic-1}
    \sigma^{(0)}(k) &= g^2 [m_V^2 + 2m_e^2]\times \frac{2}{3 m_e\Lambda}
    \frac{1}{\vb{k}^2} \times \frac{1}{\qty((a-b)^2+1)^3}~,\\
    \label{analytic-2}
    \sigma^{(1)}(k) &= g^2 [m_V^2 + 2m_e^2]\times \frac{2}{3 m_e\Lambda}
    \frac{1}{\vb{k}^2} \times \frac{\beta}{2}\mathcal{I}(a,b)~,
\end{align}
where $\vb{k}^2$ is the momentum of the incident positron,  
\begin{align}
    \!\!\!\mathcal{I}(a,b) \! = \!\int_0^1\!\! \dd x \!&  \qty[\frac{1/x^2}{\qty[(a-\frac{b}{x})^2+1]^3}- \frac{1}{\qty[(a-b)^2+1]^3}] \nonumber \\
    &\hspace{0.375\linewidth}\times\frac{1 + x^2}{1-x }~. 
\end{align}
The integral $\mathcal{I}(a,b)$ can be obtained analytically, but is lengthy. The explicit expression can be obtained with symbolic integration software or found in the \texttt{PETITE} source code {{\href{https://github.com/kjkellyphys/PETITE}{\large\color{BlueViolet}\faGithub}} . One can check that in the $\Lambda \rightarrow 0$ limit, the result approaches that for a free electron obtained with a fixed-order splitting function
\begin{equation}
    \lim_{\Lambda \rightarrow 0} \sigma^{(1)} = \frac{\beta}{2}   \frac{(2\pi)\qty[g^2 (m_V^2+2m_e^2)]}{4 \sqrt{m_V^2(m_V^2-4m_e^2)}}  \frac{1+(m_V^2/s)^2}{s-m_V^2}~.
\end{equation}
\subsection{Comparison with literature \label{sec:compare-lit} }
As mentioned above \cref{sigma-def-atomic-binding} and in \cref{footnote-binding}, our results differ (somewhat) from previous treatments in the literature \cite{Celentano:2020vtu,Marsicano:2018glj,Marsicano:2018glj,Nardi:2018cxi,Marsicano:2018krp,Arias-Aragon:2024qji}. This is because our starting point for the calculation of the cross section is \cref{atom-phase-space}, as discussed in detail in Section II of Ref.~\cite{Plestid:2024xzh}.
In previous treatments, the cross section has been modelled by averaging over the momentum distribution of an atom (see e.g., Eq.~(1) of Ref.~\cite{Arias-Aragon:2024qji}
\begin{equation}
    \label{ansatz}
    \sigma \sim \int \frac{\dd^3 p}{(2\pi)^3} n(\vb{p}) \sigma_{\rm free}(\vb{p})~, 
\end{equation}
where $\sigma_{\rm free}(\vb{p})$ is calculated on a free electron with momentum $\vb{p}$ in the lab-frame. 

As discussed in Ref.~\cite{Plestid:2024xzh} the ansatz in \cref{ansatz} does not capture bound-state effects beginning at $O(p^2/m_e^2)$. Nevertheless, as discussed above, we find that it does a good job of modelling $O(1)$ effects for reactions with kinematic thresholds and singularities (e.g., resonances). The proper formula for the cross section in terms of $n(\vb{p})$ is discussed in the main text beneath \cref{sigma-def-atomic-binding}. 

Before proceeding we will outline the advantages of our simplified treatment of the atomic momentum distribution (modelling it as a sum of $1s$ orbitals). The use of a 
 simple $1s$ wavefunction permits for the inclusion of initial state radiation in a unified framework as discussed in \cref{sec:rad-tail}, and results in the analytic expressions in \cref{analytic-1,analytic-2,analytic-3}. The ability to have analytic results is a major asset for a Monte Carlo tool like \texttt{PETITE}. If one wanted to incorporate a realistic momentum distribution $n(\vb{p})$, as advocated in Ref.~\cite{Arias-Aragon:2024qji}, then the appropriate prescription is to set  $\epsilon=0$ in all matrix elements, and replace $|\psi_A(\vb{p})|^2\rightarrow n(\vb{p})$; this is discussed beneath \cref{sigma-def-atomic-binding} in the main text. In this case the inclusion of initial state radiation would likely demand additional numerical integrations.

\subsection{Dark Compton scattering \label{sec:dark-compton-splitting}}
We can apply the methods of the previous section to calculate bound-state corrections to dark Compton scattering. This can be obtained using a splitting function formalism by folding $\beta P_{\gamma \rightarrow ee}(x)/4$, with the photon splitting function $P_{\gamma \rightarrow ee}(x) = x^2+(1-x)^2$, against the resonant cross section. Doing so one finds, 
\begin{equation}
    \label{analytic-3}
    \sigma = g^2 [m_V^2 + 2m_e^2]\times \frac{2}{3 m_e\Lambda}
    \frac{1}{\vb{k}^2} \times \frac{\beta}{4}\mathcal{J}(a,b)~,   
\end{equation}
with $a,\; b$ given in \cref{eq:ab_coef} and
\begin{equation}
    \mathcal{J}(a,b)= \int_0^1 \dd x \frac{1/x^2}{\qty[(a-\frac{b}{x})^2+1]^3} \qty( x^2+(1-x)^2)~. 
\end{equation}
The parton-level interpretation is that the photon fluctuates into a $e^+e^-$ pair with the positron annihilating on the atomic electron, while the high energy $e^-$ is emitted as a final state particle.

\section{Geometric Efficiency \label{app:Efficiency}}

\begin{figure}[hb]
    \centering
    \includegraphics[width=0.88\linewidth]{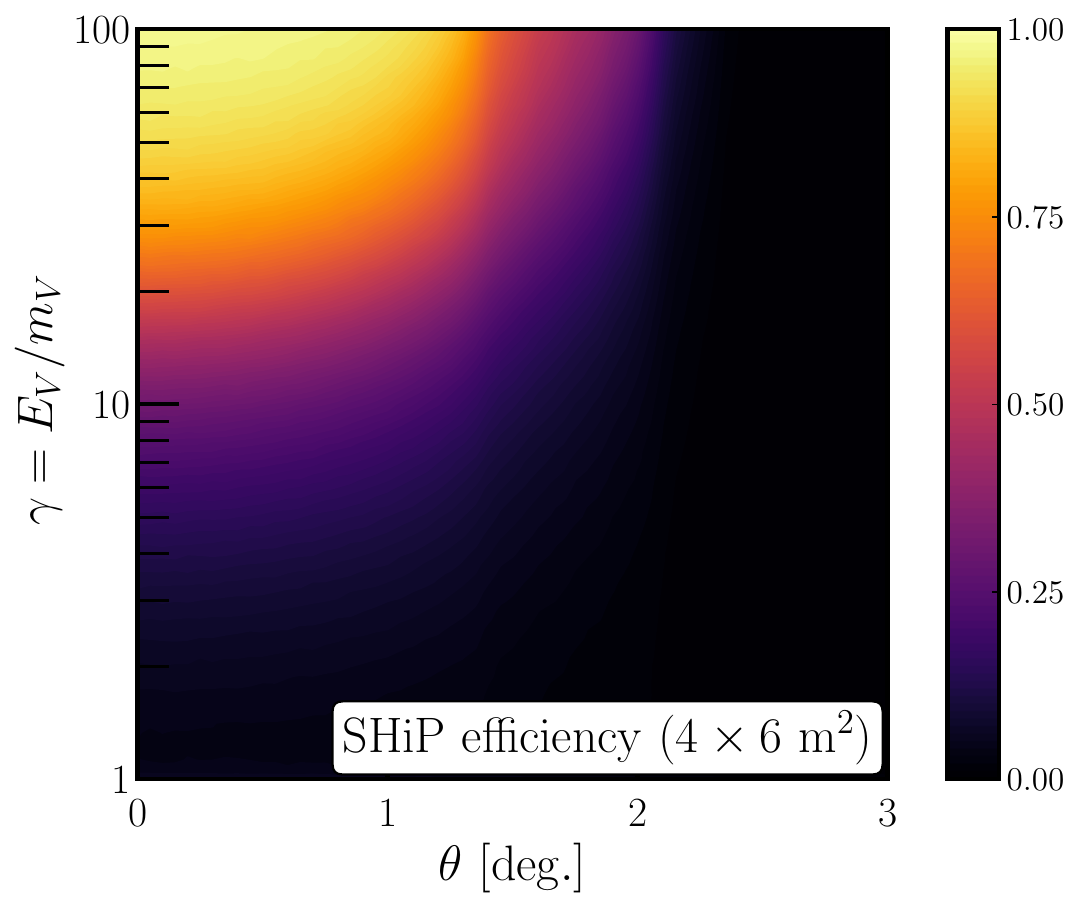}
    \caption{Geometric efficiency of the SHiP detector as a function of the dark vector's angle of emission relative to the beam axis, and the boost of the dark vector. We assume that $\lambda_V \gg 100~{\rm m}$ such that the decays are uniformly distributed in the decay pipe. \label{fig:efficiency}}
\end{figure}

\begin{figure}[ht]
    \centering
    \includegraphics[width=0.88\linewidth]{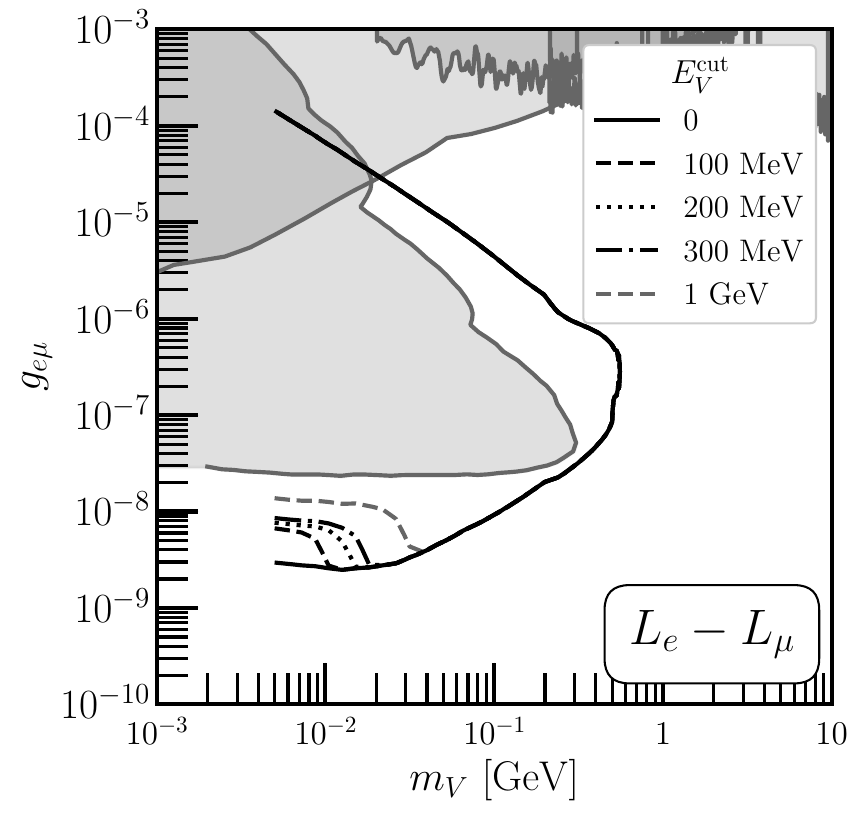}
    \caption{Impact of energy cuts, $E_V\geq E_V^{\rm cut}$ on the projected sensitivity for the $L_e - L_\mu$ leptophilic gauge boson scenario. Existing constraints (gray shaded regions) are the same as shown in~\cref{fig:SHiP_projections}.\label{fig:energy_cut}}
\end{figure}

This angular acceptance is determined by an interplay between how far along the decay pipe the dark vector has travelled before it decays and the typical direction the electrons travel, relative to the vector, after the decay. The former is determined by the lifetime, boost, and the emission angle with respect to the beam (for simplicity we assume all dark vectors are made promptly at the target) while the latter is determined solely by the boost (we assume isotropic decays in the vector's rest frame). Rather than accept/reject on an event-by-event basis we make use of the azimuthal symmetry of the electromagnetic cascade to determine an azimuthally averaged acceptance for each dark vector produced. This weight is then applied to each event to convert a flux of dark vectors $\dd \Phi_V(E_V)/\dd E_V$ into an event rate. We assume that $m_V \gg 2m_e$ so that the outgoing direction of the electron/positron in the lab frame are determined purely by the vector's boost and direction.

\begin{figure*}[ht]
    \centering
    \includegraphics[width=\textwidth]{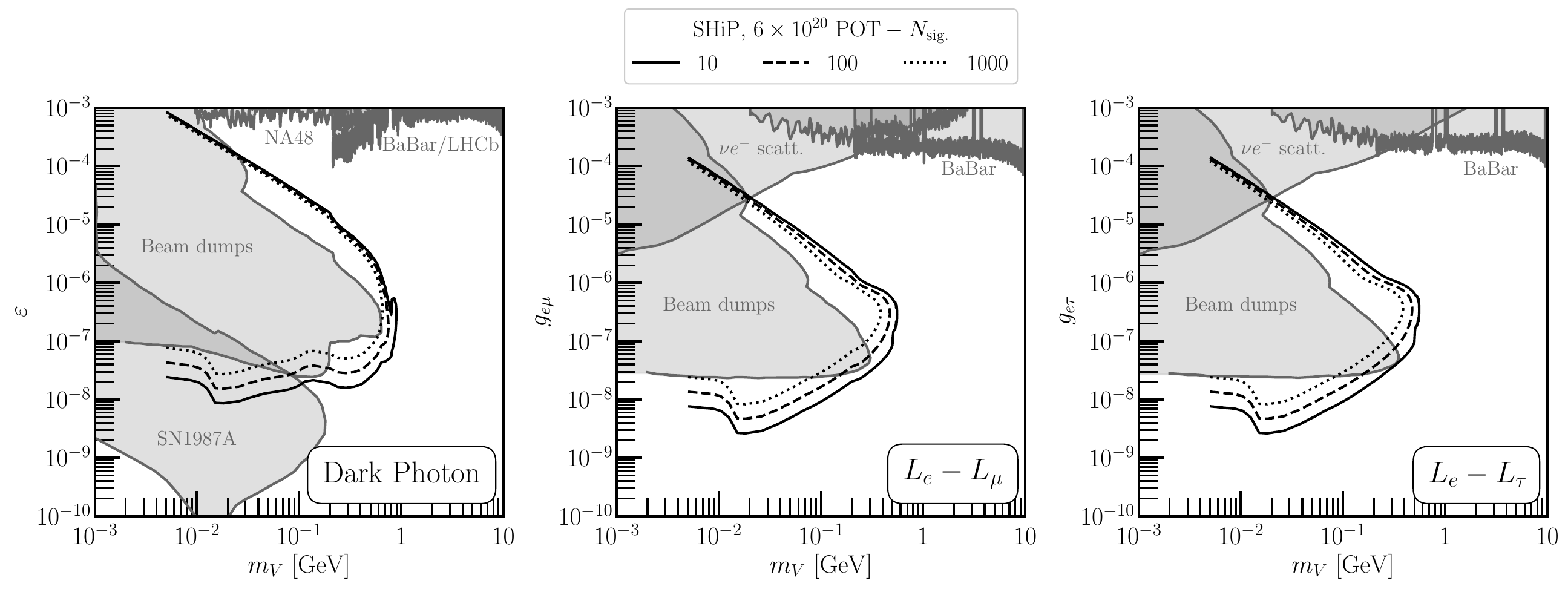}
    \caption{Impact of signal rate requirements on the projected sensitivity at SHiP, assuming $6\times 10^{20}$ protons on target.  Existing constraints (gray shaded regions) are the same as shown in~\cref{fig:SHiP_projections}.\label{fig:sensitivityvaryNsig}}
\end{figure*}

The weight is the product of the probability for the dark vector to decay before the detector and the (azimuthally averaged) likelihood that both the decay products reach the detector, the latter of which we dub the ``geometric efficiency''. 
To determine the geometric efficiency we perform a Monte Carlo simulation which decays vectors, of varying boosts $\gamma_V$, at different locations along the beam pipe. All vectors are taken to propagate in the $x-z$ plane so that the efficiency is a function of angle relative to beam axis, $\theta_V$, and distance from the target, $r_V$. The decay takes place isotropically in the vector's rest frame and is boosted back to the lab frame, and to account for the non-symmetric detector we uniformly rotate the detector about the beam axis to trace out an azimuthally averaged profile. This is what is effectively ``seen" by showers. Once this geometric efficiency is known, as a function of $\gamma_V$, $r_V$, and $\theta_V$, it can be convoluted with the probability for a dark vector in a shower to decay after propagating distance $r_V$, given in \cref{eq:Nobsscaling}.

We show the resulting efficiency in \cref{fig:efficiency}, plotted in the $\theta_V-\gamma_V$ plane. Here we have assumed the lab-frame decay length of the vector is large, $\lambda_V\gg 100$~m, which means the convolution is essentially flat in $r_V$ and we have normalised away the probability to decay before the detector so that what is shown is solely the geometric efficiency. We find that for $\theta_V\lesssim 1^\circ$, once $\gamma \gtrsim 3$ the geometric efficiency is $\sim 0.25$ while for $\gamma \gtrsim 10$ it quickly approaches unity. The geometric efficiency (slightly) reduces the sensitivity gains at low masses where dark vectors produced from resonant annihilation tend to have smaller boosts. 

\section{Effect of energy thresholds \label{app:energythreshold}}

In \cref{fig:energy_cut} we show the impact of including energy cuts at SHiP. The main effect is to sculpt the resonant annihilation sensitivity. Since the peak of the annihilation flux lies at 
\begin{equation}
E_V^{\rm res}\simeq 100\; \MeV \left(\frac{m_V}{10\;\MeV}\right)^2   ~, 
\end{equation}
the sensitivity is substantially affected by energy cuts with $E_V^{\rm cut} \geq E^{\rm res}_V$. 

As can be seen in  \cref{fig:energy_cut}, provided $E_V^{\rm cut}\leq 300~{\rm MeV}$, the sensitivity to dark vectors with mass $m_V\geq 20~{\rm MeV}$ remains excellent. Based on conversations with the SHiP collaboration \cite{Nico:chat} we are optimistic that this is possible and with dedicated studies it may even be possible to go as low as $E_V^{\rm cut}=100\,{\rm MeV}$, which lowers the point at which sensitivity is impacted by a further factor of $2$. Our results call for a systematic investigation into low-energy (i.e., sub-GeV energy) backgrounds and detection prospects. 

\vfill 
\pagebreak 

\section{Sensitivity for different signal rate requirements\label{app:varyingNsig}}

In \cref{fig:SHiP_projections} we assumed the search for an electron-positron reconstructing the dark vector mass was relatively background free and thus ten signal events, $N_{\mathrm{sig.}}=10$, was sufficient to determine sensitivity at SHiP.  However, if there are some as-yet unidentified backgrounds the required number of signal events may need to be increased.  In \cref{fig:sensitivityvaryNsig} we show the effect on sensitivity for $N_{\mathrm{sig.}}=100, 1000$.

\vfill
\pagebreak

~
\pagebreak

\bibliographystyle{apsrev4-1}
\bibliography{refs}

\begin{thebibliography}{92}%
\makeatletter
\providecommand \@ifxundefined [1]{%
 \@ifx{#1\undefined}
}%
\providecommand \@ifnum [1]{%
 \ifnum #1\expandafter \@firstoftwo
 \else \expandafter \@secondoftwo
 \fi
}%
\providecommand \@ifx [1]{%
 \ifx #1\expandafter \@firstoftwo
 \else \expandafter \@secondoftwo
 \fi
}%
\providecommand \natexlab [1]{#1}%
\providecommand \enquote  [1]{``#1''}%
\providecommand \bibnamefont  [1]{#1}%
\providecommand \bibfnamefont [1]{#1}%
\providecommand \citenamefont [1]{#1}%
\providecommand \href@noop [0]{\@secondoftwo}%
\providecommand \href [0]{\begingroup \@sanitize@url \@href}%
\providecommand \@href[1]{\@@startlink{#1}\@@href}%
\providecommand \@@href[1]{\endgroup#1\@@endlink}%
\providecommand \@sanitize@url [0]{\catcode `\\12\catcode `\$12\catcode `\&12\catcode `\#12\catcode `\^12\catcode `\_12\catcode `\%12\relax}%
\providecommand \@@startlink[1]{}%
\providecommand \@@endlink[0]{}%
\providecommand \url  [0]{\begingroup\@sanitize@url \@url }%
\providecommand \@url [1]{\endgroup\@href {#1}{\urlprefix }}%
\providecommand \urlprefix  [0]{URL }%
\providecommand \Eprint [0]{\href }%
\providecommand \doibase [0]{http://dx.doi.org/}%
\providecommand \selectlanguage [0]{\@gobble}%
\providecommand \bibinfo  [0]{\@secondoftwo}%
\providecommand \bibfield  [0]{\@secondoftwo}%
\providecommand \translation [1]{[#1]}%
\providecommand \BibitemOpen [0]{}%
\providecommand \bibitemStop [0]{}%
\providecommand \bibitemNoStop [0]{.\EOS\space}%
\providecommand \EOS [0]{\spacefactor3000\relax}%
\providecommand \BibitemShut  [1]{\csname bibitem#1\endcsname}%
\let\auto@bib@innerbib\@empty
\bibitem [{\citenamefont {Artuso}\ \emph {et~al.}(2022)\citenamefont {Artuso} \emph {et~al.}}]{Artuso:2022ouk}%
  \BibitemOpen
  \bibfield  {author} {\bibinfo {author} {\bibfnamefont {M.}~\bibnamefont {Artuso}} \emph {et~al.},\ }\href@noop {} {\  (\bibinfo {year} {2022})},\ \Eprint {http://arxiv.org/abs/2210.04765} {arXiv:2210.04765 [hep-ex]} \BibitemShut {NoStop}%
\bibitem [{\citenamefont {Antel}\ \emph {et~al.}(2023)\citenamefont {Antel} \emph {et~al.}}]{Antel:2023hkf}%
  \BibitemOpen
  \bibfield  {author} {\bibinfo {author} {\bibfnamefont {C.}~\bibnamefont {Antel}} \emph {et~al.},\ }\href {\doibase 10.1140/epjc/s10052-023-12168-5} {\bibfield  {journal} {\bibinfo  {journal} {Eur. Phys. J. C}\ }\textbf {\bibinfo {volume} {83}},\ \bibinfo {pages} {1122} (\bibinfo {year} {2023})},\ \Eprint {http://arxiv.org/abs/2305.01715} {arXiv:2305.01715 [hep-ph]} \BibitemShut {NoStop}%
\bibitem [{\citenamefont {Baxter}\ \emph {et~al.}(2020)\citenamefont {Baxter} \emph {et~al.}}]{Baxter:2019mcx}%
  \BibitemOpen
  \bibfield  {author} {\bibinfo {author} {\bibfnamefont {D.}~\bibnamefont {Baxter}} \emph {et~al.},\ }\href {\doibase 10.1007/JHEP02(2020)123} {\bibfield  {journal} {\bibinfo  {journal} {JHEP}\ }\textbf {\bibinfo {volume} {02}},\ \bibinfo {pages} {123} (\bibinfo {year} {2020})},\ \Eprint {http://arxiv.org/abs/1911.00762} {arXiv:1911.00762 [physics.ins-det]} \BibitemShut {NoStop}%
\bibitem [{\citenamefont {Barbeau}\ \emph {et~al.}(2023)\citenamefont {Barbeau}, \citenamefont {Efremenko},\ and\ \citenamefont {Scholberg}}]{Barbeau:2021exu}%
  \BibitemOpen
  \bibfield  {author} {\bibinfo {author} {\bibfnamefont {P.~S.}\ \bibnamefont {Barbeau}}, \bibinfo {author} {\bibfnamefont {Y.}~\bibnamefont {Efremenko}}, \ and\ \bibinfo {author} {\bibfnamefont {K.}~\bibnamefont {Scholberg}},\ }\href {\doibase 10.1146/annurev-nucl-101918-023518} {\bibfield  {journal} {\bibinfo  {journal} {Ann. Rev. Nucl. Part. Sci.}\ }\textbf {\bibinfo {volume} {73}},\ \bibinfo {pages} {41} (\bibinfo {year} {2023})},\ \Eprint {http://arxiv.org/abs/2111.07033} {arXiv:2111.07033 [hep-ex]} \BibitemShut {NoStop}%
\bibitem [{\citenamefont {Ajimura}\ \emph {et~al.}(2021)\citenamefont {Ajimura} \emph {et~al.}}]{JSNS2:2021hyk}%
  \BibitemOpen
  \bibfield  {author} {\bibinfo {author} {\bibfnamefont {S.}~\bibnamefont {Ajimura}} \emph {et~al.} (\bibinfo {collaboration} {JSNS2}),\ }\href {\doibase 10.1016/j.nima.2021.165742} {\bibfield  {journal} {\bibinfo  {journal} {Nucl. Instrum. Meth. A}\ }\textbf {\bibinfo {volume} {1014}},\ \bibinfo {pages} {165742} (\bibinfo {year} {2021})},\ \Eprint {http://arxiv.org/abs/2104.13169} {arXiv:2104.13169 [physics.ins-det]} \BibitemShut {NoStop}%
\bibitem [{\citenamefont {Aguilar-Arevalo}\ \emph {et~al.}(2022)\citenamefont {Aguilar-Arevalo} \emph {et~al.}}]{CCM:2021leg}%
  \BibitemOpen
  \bibfield  {author} {\bibinfo {author} {\bibfnamefont {A.~A.}\ \bibnamefont {Aguilar-Arevalo}} \emph {et~al.} (\bibinfo {collaboration} {CCM}),\ }\href {\doibase 10.1103/PhysRevD.106.012001} {\bibfield  {journal} {\bibinfo  {journal} {Phys. Rev. D}\ }\textbf {\bibinfo {volume} {106}},\ \bibinfo {pages} {012001} (\bibinfo {year} {2022})},\ \Eprint {http://arxiv.org/abs/2105.14020} {arXiv:2105.14020 [hep-ex]} \BibitemShut {NoStop}%
\bibitem [{\citenamefont {Aguilar-Arevalo}\ \emph {et~al.}(2023)\citenamefont {Aguilar-Arevalo} \emph {et~al.}}]{Aguilar-Arevalo:2023dai}%
  \BibitemOpen
  \bibfield  {author} {\bibinfo {author} {\bibfnamefont {A.~A.}\ \bibnamefont {Aguilar-Arevalo}} \emph {et~al.},\ }\href@noop {} {\  (\bibinfo {year} {2023})},\ \Eprint {http://arxiv.org/abs/2311.09915} {arXiv:2311.09915 [hep-ex]} \BibitemShut {NoStop}%
\bibitem [{\citenamefont {Abe}\ \emph {et~al.}(2011)\citenamefont {Abe} \emph {et~al.}}]{T2K:2011qtm}%
  \BibitemOpen
  \bibfield  {author} {\bibinfo {author} {\bibfnamefont {K.}~\bibnamefont {Abe}} \emph {et~al.} (\bibinfo {collaboration} {T2K}),\ }\href {\doibase 10.1016/j.nima.2011.06.067} {\bibfield  {journal} {\bibinfo  {journal} {Nucl. Instrum. Meth. A}\ }\textbf {\bibinfo {volume} {659}},\ \bibinfo {pages} {106} (\bibinfo {year} {2011})},\ \Eprint {http://arxiv.org/abs/1106.1238} {arXiv:1106.1238 [physics.ins-det]} \BibitemShut {NoStop}%
\bibitem [{\citenamefont {Machado}\ \emph {et~al.}(2019)\citenamefont {Machado}, \citenamefont {Palamara},\ and\ \citenamefont {Schmitz}}]{Machado:2019oxb}%
  \BibitemOpen
  \bibfield  {author} {\bibinfo {author} {\bibfnamefont {P.~A.}\ \bibnamefont {Machado}}, \bibinfo {author} {\bibfnamefont {O.}~\bibnamefont {Palamara}}, \ and\ \bibinfo {author} {\bibfnamefont {D.~W.}\ \bibnamefont {Schmitz}},\ }\href {\doibase 10.1146/annurev-nucl-101917-020949} {\bibfield  {journal} {\bibinfo  {journal} {Ann. Rev. Nucl. Part. Sci.}\ }\textbf {\bibinfo {volume} {69}},\ \bibinfo {pages} {363} (\bibinfo {year} {2019})},\ \Eprint {http://arxiv.org/abs/1903.04608} {arXiv:1903.04608 [hep-ex]} \BibitemShut {NoStop}%
\bibitem [{\citenamefont {Acciarri}\ \emph {et~al.}(2023)\citenamefont {Acciarri} \emph {et~al.}}]{ArgoNeuT:2022mrm}%
  \BibitemOpen
  \bibfield  {author} {\bibinfo {author} {\bibfnamefont {R.}~\bibnamefont {Acciarri}} \emph {et~al.} (\bibinfo {collaboration} {ArgoNeuT}),\ }\href {\doibase 10.1103/PhysRevLett.130.221802} {\bibfield  {journal} {\bibinfo  {journal} {Phys. Rev. Lett.}\ }\textbf {\bibinfo {volume} {130}},\ \bibinfo {pages} {221802} (\bibinfo {year} {2023})},\ \Eprint {http://arxiv.org/abs/2207.08448} {arXiv:2207.08448 [hep-ex]} \BibitemShut {NoStop}%
\bibitem [{\citenamefont {Abed~Abud}\ \emph {et~al.}(2022)\citenamefont {Abed~Abud} \emph {et~al.}}]{DUNE:2022aul}%
  \BibitemOpen
  \bibfield  {author} {\bibinfo {author} {\bibfnamefont {A.}~\bibnamefont {Abed~Abud}} \emph {et~al.} (\bibinfo {collaboration} {DUNE}),\ }\href@noop {} {\  (\bibinfo {year} {2022})},\ \Eprint {http://arxiv.org/abs/2203.06100} {arXiv:2203.06100 [hep-ex]} \BibitemShut {NoStop}%
\bibitem [{\citenamefont {Banerjee}\ \emph {et~al.}(2019)\citenamefont {Banerjee} \emph {et~al.}}]{Banerjee:2019pds}%
  \BibitemOpen
  \bibfield  {author} {\bibinfo {author} {\bibfnamefont {D.}~\bibnamefont {Banerjee}} \emph {et~al.},\ }\href {\doibase 10.1103/PhysRevLett.123.121801} {\bibfield  {journal} {\bibinfo  {journal} {Phys. Rev. Lett.}\ }\textbf {\bibinfo {volume} {123}},\ \bibinfo {pages} {121801} (\bibinfo {year} {2019})},\ \Eprint {http://arxiv.org/abs/1906.00176} {arXiv:1906.00176 [hep-ex]} \BibitemShut {NoStop}%
\bibitem [{\citenamefont {\r{A}kesson}\ \emph {et~al.}(2018)\citenamefont {\r{A}kesson} \emph {et~al.}}]{LDMX:2018cma}%
  \BibitemOpen
  \bibfield  {author} {\bibinfo {author} {\bibfnamefont {T.}~\bibnamefont {\r{A}kesson}} \emph {et~al.} (\bibinfo {collaboration} {LDMX}),\ }\href@noop {} {\  (\bibinfo {year} {2018})},\ \Eprint {http://arxiv.org/abs/1808.05219} {arXiv:1808.05219 [hep-ex]} \BibitemShut {NoStop}%
\bibitem [{\citenamefont {Battaglieri}\ \emph {et~al.}(2022)\citenamefont {Battaglieri} \emph {et~al.}}]{Battaglieri:2022dcy}%
  \BibitemOpen
  \bibfield  {author} {\bibinfo {author} {\bibfnamefont {M.}~\bibnamefont {Battaglieri}} \emph {et~al.},\ }\href {\doibase 10.1103/PhysRevD.106.072011} {\bibfield  {journal} {\bibinfo  {journal} {Phys. Rev. D}\ }\textbf {\bibinfo {volume} {106}},\ \bibinfo {pages} {072011} (\bibinfo {year} {2022})},\ \Eprint {http://arxiv.org/abs/2208.01387} {arXiv:2208.01387 [hep-ex]} \BibitemShut {NoStop}%
\bibitem [{\citenamefont {Ahdida}\ \emph {et~al.}(2022)\citenamefont {Ahdida} \emph {et~al.}}]{SHiP:2021nfo}%
  \BibitemOpen
  \bibfield  {author} {\bibinfo {author} {\bibfnamefont {C.}~\bibnamefont {Ahdida}} \emph {et~al.} (\bibinfo {collaboration} {SHiP}),\ }\href {\doibase 10.1140/epjc/s10052-022-10346-5} {\bibfield  {journal} {\bibinfo  {journal} {Eur. Phys. J. C}\ }\textbf {\bibinfo {volume} {82}},\ \bibinfo {pages} {486} (\bibinfo {year} {2022})},\ \Eprint {http://arxiv.org/abs/2112.01487} {arXiv:2112.01487 [physics.ins-det]} \BibitemShut {NoStop}%
\bibitem [{\citenamefont {Aberle}\ \emph {et~al.}(2022)\citenamefont {Aberle} \emph {et~al.}}]{Aberle:2839677}%
  \BibitemOpen
  \bibfield  {author} {\bibinfo {author} {\bibfnamefont {O.}~\bibnamefont {Aberle}} \emph {et~al.} (\bibinfo {collaboration} {SHiP}),\ }\href {https://cds.cern.ch/record/2839677} {\emph {\bibinfo {title} {{BDF/SHiP at the ECN3 high-intensity beam facility}}}},\ \bibinfo {type} {Tech. Rep.}\ (\bibinfo  {institution} {CERN},\ \bibinfo {address} {Geneva},\ \bibinfo {year} {2022})\BibitemShut {NoStop}%
\bibitem [{\citenamefont {Albanese}\ \emph {et~al.}(2023)\citenamefont {Albanese} \emph {et~al.}}]{Albanese:2878604}%
  \BibitemOpen
  \bibfield  {author} {\bibinfo {author} {\bibfnamefont {R.}~\bibnamefont {Albanese}} \emph {et~al.} (\bibinfo {collaboration} {SHiP}),\ }\href {https://cds.cern.ch/record/2878604} {\emph {\bibinfo {title} {{BDF/SHiP at the ECN3 high-intensity beam facility}}}},\ \bibinfo {type} {Tech. Rep.}\ (\bibinfo  {institution} {CERN},\ \bibinfo {address} {Geneva},\ \bibinfo {year} {2023})\BibitemShut {NoStop}%
\bibitem [{\citenamefont {Beacham}\ \emph {et~al.}(2020)\citenamefont {Beacham} \emph {et~al.}}]{Beacham:2019nyx}%
  \BibitemOpen
  \bibfield  {author} {\bibinfo {author} {\bibfnamefont {J.}~\bibnamefont {Beacham}} \emph {et~al.},\ }\href {\doibase 10.1088/1361-6471/ab4cd2} {\bibfield  {journal} {\bibinfo  {journal} {J. Phys. G}\ }\textbf {\bibinfo {volume} {47}},\ \bibinfo {pages} {010501} (\bibinfo {year} {2020})},\ \Eprint {http://arxiv.org/abs/1901.09966} {arXiv:1901.09966 [hep-ex]} \BibitemShut {NoStop}%
\bibitem [{\citenamefont {Ilten}\ \emph {et~al.}(2022)\citenamefont {Ilten} \emph {et~al.}}]{Ilten:2022lfq}%
  \BibitemOpen
  \bibfield  {author} {\bibinfo {author} {\bibfnamefont {P.}~\bibnamefont {Ilten}} \emph {et~al.},\ }in\ \href@noop {} {\emph {\bibinfo {booktitle} {{Snowmass 2021}}}}\ (\bibinfo {year} {2022})\ \Eprint {http://arxiv.org/abs/2206.04220} {arXiv:2206.04220 [hep-ex]} \BibitemShut {NoStop}%
\bibitem [{\citenamefont {Donnelly}\ \emph {et~al.}(1978)\citenamefont {Donnelly}, \citenamefont {Freedman}, \citenamefont {Lytel}, \citenamefont {Peccei},\ and\ \citenamefont {Schwartz}}]{Donnelly:1978ty}%
  \BibitemOpen
  \bibfield  {author} {\bibinfo {author} {\bibfnamefont {T.~W.}\ \bibnamefont {Donnelly}}, \bibinfo {author} {\bibfnamefont {S.~J.}\ \bibnamefont {Freedman}}, \bibinfo {author} {\bibfnamefont {R.~S.}\ \bibnamefont {Lytel}}, \bibinfo {author} {\bibfnamefont {R.~D.}\ \bibnamefont {Peccei}}, \ and\ \bibinfo {author} {\bibfnamefont {M.}~\bibnamefont {Schwartz}},\ }\href {\doibase 10.1103/PhysRevD.18.1607} {\bibfield  {journal} {\bibinfo  {journal} {Phys. Rev. D}\ }\textbf {\bibinfo {volume} {18}},\ \bibinfo {pages} {1607} (\bibinfo {year} {1978})}\BibitemShut {NoStop}%
\bibitem [{\citenamefont {Tsai}(1986)}]{Tsai:1986tx}%
  \BibitemOpen
  \bibfield  {author} {\bibinfo {author} {\bibfnamefont {Y.-S.}\ \bibnamefont {Tsai}},\ }\href {\doibase 10.1103/PhysRevD.34.1326} {\bibfield  {journal} {\bibinfo  {journal} {Phys. Rev. D}\ }\textbf {\bibinfo {volume} {34}},\ \bibinfo {pages} {1326} (\bibinfo {year} {1986})}\BibitemShut {NoStop}%
\bibitem [{\citenamefont {Bjorken}\ \emph {et~al.}(1988)\citenamefont {Bjorken}, \citenamefont {Ecklund}, \citenamefont {Nelson}, \citenamefont {Abashian}, \citenamefont {Church}, \citenamefont {Lu}, \citenamefont {Mo}, \citenamefont {Nunamaker},\ and\ \citenamefont {Rassmann}}]{Bjorken:1988as}%
  \BibitemOpen
  \bibfield  {author} {\bibinfo {author} {\bibfnamefont {J.~D.}\ \bibnamefont {Bjorken}}, \bibinfo {author} {\bibfnamefont {S.}~\bibnamefont {Ecklund}}, \bibinfo {author} {\bibfnamefont {W.~R.}\ \bibnamefont {Nelson}}, \bibinfo {author} {\bibfnamefont {A.}~\bibnamefont {Abashian}}, \bibinfo {author} {\bibfnamefont {C.}~\bibnamefont {Church}}, \bibinfo {author} {\bibfnamefont {B.}~\bibnamefont {Lu}}, \bibinfo {author} {\bibfnamefont {L.~W.}\ \bibnamefont {Mo}}, \bibinfo {author} {\bibfnamefont {T.~A.}\ \bibnamefont {Nunamaker}}, \ and\ \bibinfo {author} {\bibfnamefont {P.}~\bibnamefont {Rassmann}},\ }\href {\doibase 10.1103/PhysRevD.38.3375} {\bibfield  {journal} {\bibinfo  {journal} {Phys. Rev. D}\ }\textbf {\bibinfo {volume} {38}},\ \bibinfo {pages} {3375} (\bibinfo {year} {1988})}\BibitemShut {NoStop}%
\bibitem [{\citenamefont {Prinz}\ \emph {et~al.}(1998)\citenamefont {Prinz} \emph {et~al.}}]{Prinz:1998ua}%
  \BibitemOpen
  \bibfield  {author} {\bibinfo {author} {\bibfnamefont {A.~A.}\ \bibnamefont {Prinz}} \emph {et~al.},\ }\href {\doibase 10.1103/PhysRevLett.81.1175} {\bibfield  {journal} {\bibinfo  {journal} {Phys. Rev. Lett.}\ }\textbf {\bibinfo {volume} {81}},\ \bibinfo {pages} {1175} (\bibinfo {year} {1998})},\ \Eprint {http://arxiv.org/abs/hep-ex/9804008} {arXiv:hep-ex/9804008} \BibitemShut {NoStop}%
\bibitem [{\citenamefont {Prinz}(2001)}]{Prinz:2001qz}%
  \BibitemOpen
  \bibfield  {author} {\bibinfo {author} {\bibfnamefont {A.~A.}\ \bibnamefont {Prinz}},\ }\emph {\bibinfo {title} {{The Search for millicharged particles at SLAC}}},\ \href@noop {} {\bibinfo {type} {Other thesis}},\ \bibinfo  {school} {Stanford University} (\bibinfo {year} {2001})\BibitemShut {NoStop}%
\bibitem [{\citenamefont {Marsicano}\ \emph {et~al.}(2018{\natexlab{a}})\citenamefont {Marsicano}, \citenamefont {Battaglieri}, \citenamefont {Bond\'\i{}}, \citenamefont {Carvajal}, \citenamefont {Celentano}, \citenamefont {De~Napoli}, \citenamefont {De~Vita}, \citenamefont {Nardi}, \citenamefont {Raggi},\ and\ \citenamefont {Valente}}]{Marsicano:2018glj}%
  \BibitemOpen
  \bibfield  {author} {\bibinfo {author} {\bibfnamefont {L.}~\bibnamefont {Marsicano}}, \bibinfo {author} {\bibfnamefont {M.}~\bibnamefont {Battaglieri}}, \bibinfo {author} {\bibfnamefont {M.}~\bibnamefont {Bond\'\i{}}}, \bibinfo {author} {\bibfnamefont {C.~D.~R.}\ \bibnamefont {Carvajal}}, \bibinfo {author} {\bibfnamefont {A.}~\bibnamefont {Celentano}}, \bibinfo {author} {\bibfnamefont {M.}~\bibnamefont {De~Napoli}}, \bibinfo {author} {\bibfnamefont {R.}~\bibnamefont {De~Vita}}, \bibinfo {author} {\bibfnamefont {E.}~\bibnamefont {Nardi}}, \bibinfo {author} {\bibfnamefont {M.}~\bibnamefont {Raggi}}, \ and\ \bibinfo {author} {\bibfnamefont {P.}~\bibnamefont {Valente}},\ }\href {\doibase 10.1103/PhysRevLett.121.041802} {\bibfield  {journal} {\bibinfo  {journal} {Phys. Rev. Lett.}\ }\textbf {\bibinfo {volume} {121}},\ \bibinfo {pages} {041802} (\bibinfo {year} {2018}{\natexlab{a}})},\ \Eprint {http://arxiv.org/abs/1807.05884} {arXiv:1807.05884 [hep-ex]} \BibitemShut {NoStop}%
\bibitem [{\citenamefont {Nardi}\ \emph {et~al.}(2018)\citenamefont {Nardi}, \citenamefont {Carvajal}, \citenamefont {Ghoshal}, \citenamefont {Meloni},\ and\ \citenamefont {Raggi}}]{Nardi:2018cxi}%
  \BibitemOpen
  \bibfield  {author} {\bibinfo {author} {\bibfnamefont {E.}~\bibnamefont {Nardi}}, \bibinfo {author} {\bibfnamefont {C.~D.~R.}\ \bibnamefont {Carvajal}}, \bibinfo {author} {\bibfnamefont {A.}~\bibnamefont {Ghoshal}}, \bibinfo {author} {\bibfnamefont {D.}~\bibnamefont {Meloni}}, \ and\ \bibinfo {author} {\bibfnamefont {M.}~\bibnamefont {Raggi}},\ }\href {\doibase 10.1103/PhysRevD.97.095004} {\bibfield  {journal} {\bibinfo  {journal} {Phys. Rev. D}\ }\textbf {\bibinfo {volume} {97}},\ \bibinfo {pages} {095004} (\bibinfo {year} {2018})},\ \Eprint {http://arxiv.org/abs/1802.04756} {arXiv:1802.04756 [hep-ph]} \BibitemShut {NoStop}%
\bibitem [{\citenamefont {Marsicano}\ \emph {et~al.}(2018{\natexlab{b}})\citenamefont {Marsicano}, \citenamefont {Battaglieri}, \citenamefont {Bondi'}, \citenamefont {Carvajal}, \citenamefont {Celentano}, \citenamefont {De~Napoli}, \citenamefont {De~Vita}, \citenamefont {Nardi}, \citenamefont {Raggi},\ and\ \citenamefont {Valente}}]{Marsicano:2018krp}%
  \BibitemOpen
  \bibfield  {author} {\bibinfo {author} {\bibfnamefont {L.}~\bibnamefont {Marsicano}}, \bibinfo {author} {\bibfnamefont {M.}~\bibnamefont {Battaglieri}}, \bibinfo {author} {\bibfnamefont {M.}~\bibnamefont {Bondi'}}, \bibinfo {author} {\bibfnamefont {C.~D.~R.}\ \bibnamefont {Carvajal}}, \bibinfo {author} {\bibfnamefont {A.}~\bibnamefont {Celentano}}, \bibinfo {author} {\bibfnamefont {M.}~\bibnamefont {De~Napoli}}, \bibinfo {author} {\bibfnamefont {R.}~\bibnamefont {De~Vita}}, \bibinfo {author} {\bibfnamefont {E.}~\bibnamefont {Nardi}}, \bibinfo {author} {\bibfnamefont {M.}~\bibnamefont {Raggi}}, \ and\ \bibinfo {author} {\bibfnamefont {P.}~\bibnamefont {Valente}},\ }\href {\doibase 10.1103/PhysRevD.98.015031} {\bibfield  {journal} {\bibinfo  {journal} {Phys. Rev. D}\ }\textbf {\bibinfo {volume} {98}},\ \bibinfo {pages} {015031} (\bibinfo {year} {2018}{\natexlab{b}})},\ \Eprint {http://arxiv.org/abs/1802.03794} {arXiv:1802.03794 [hep-ex]} \BibitemShut {NoStop}%
\bibitem [{\citenamefont {Celentano}\ \emph {et~al.}(2020)\citenamefont {Celentano}, \citenamefont {Darm\'e}, \citenamefont {Marsicano},\ and\ \citenamefont {Nardi}}]{Celentano:2020vtu}%
  \BibitemOpen
  \bibfield  {author} {\bibinfo {author} {\bibfnamefont {A.}~\bibnamefont {Celentano}}, \bibinfo {author} {\bibfnamefont {L.}~\bibnamefont {Darm\'e}}, \bibinfo {author} {\bibfnamefont {L.}~\bibnamefont {Marsicano}}, \ and\ \bibinfo {author} {\bibfnamefont {E.}~\bibnamefont {Nardi}},\ }\href {\doibase 10.1103/PhysRevD.102.075026} {\bibfield  {journal} {\bibinfo  {journal} {Phys. Rev. D}\ }\textbf {\bibinfo {volume} {102}},\ \bibinfo {pages} {075026} (\bibinfo {year} {2020})},\ \Eprint {http://arxiv.org/abs/2006.09419} {arXiv:2006.09419 [hep-ph]} \BibitemShut {NoStop}%
\bibitem [{\citenamefont {Dutta}\ \emph {et~al.}(2022)\citenamefont {Dutta}, \citenamefont {Kim}, \citenamefont {Liao}, \citenamefont {Park}, \citenamefont {Shin}, \citenamefont {Strigari},\ and\ \citenamefont {Thompson}}]{Dutta:2020vop}%
  \BibitemOpen
  \bibfield  {author} {\bibinfo {author} {\bibfnamefont {B.}~\bibnamefont {Dutta}}, \bibinfo {author} {\bibfnamefont {D.}~\bibnamefont {Kim}}, \bibinfo {author} {\bibfnamefont {S.}~\bibnamefont {Liao}}, \bibinfo {author} {\bibfnamefont {J.-C.}\ \bibnamefont {Park}}, \bibinfo {author} {\bibfnamefont {S.}~\bibnamefont {Shin}}, \bibinfo {author} {\bibfnamefont {L.~E.}\ \bibnamefont {Strigari}}, \ and\ \bibinfo {author} {\bibfnamefont {A.}~\bibnamefont {Thompson}},\ }\href {\doibase 10.1007/JHEP01(2022)144} {\bibfield  {journal} {\bibinfo  {journal} {JHEP}\ }\textbf {\bibinfo {volume} {01}},\ \bibinfo {pages} {144} (\bibinfo {year} {2022})},\ \Eprint {http://arxiv.org/abs/2006.09386} {arXiv:2006.09386 [hep-ph]} \BibitemShut {NoStop}%
\bibitem [{\citenamefont {Brdar}\ \emph {et~al.}(2021)\citenamefont {Brdar}, \citenamefont {Dutta}, \citenamefont {Jang}, \citenamefont {Kim}, \citenamefont {Shoemaker}, \citenamefont {Tabrizi}, \citenamefont {Thompson},\ and\ \citenamefont {Yu}}]{Brdar:2020dpr}%
  \BibitemOpen
  \bibfield  {author} {\bibinfo {author} {\bibfnamefont {V.}~\bibnamefont {Brdar}}, \bibinfo {author} {\bibfnamefont {B.}~\bibnamefont {Dutta}}, \bibinfo {author} {\bibfnamefont {W.}~\bibnamefont {Jang}}, \bibinfo {author} {\bibfnamefont {D.}~\bibnamefont {Kim}}, \bibinfo {author} {\bibfnamefont {I.~M.}\ \bibnamefont {Shoemaker}}, \bibinfo {author} {\bibfnamefont {Z.}~\bibnamefont {Tabrizi}}, \bibinfo {author} {\bibfnamefont {A.}~\bibnamefont {Thompson}}, \ and\ \bibinfo {author} {\bibfnamefont {J.}~\bibnamefont {Yu}},\ }\href {\doibase 10.1103/PhysRevLett.126.201801} {\bibfield  {journal} {\bibinfo  {journal} {Phys. Rev. Lett.}\ }\textbf {\bibinfo {volume} {126}},\ \bibinfo {pages} {201801} (\bibinfo {year} {2021})},\ \Eprint {http://arxiv.org/abs/2011.07054} {arXiv:2011.07054 [hep-ph]} \BibitemShut {NoStop}%
\bibitem [{\citenamefont {Capozzi}\ \emph {et~al.}(2021)\citenamefont {Capozzi}, \citenamefont {Dutta}, \citenamefont {Gurung}, \citenamefont {Jang}, \citenamefont {Shoemaker}, \citenamefont {Thompson},\ and\ \citenamefont {Yu}}]{Capozzi:2021nmp}%
  \BibitemOpen
  \bibfield  {author} {\bibinfo {author} {\bibfnamefont {F.}~\bibnamefont {Capozzi}}, \bibinfo {author} {\bibfnamefont {B.}~\bibnamefont {Dutta}}, \bibinfo {author} {\bibfnamefont {G.}~\bibnamefont {Gurung}}, \bibinfo {author} {\bibfnamefont {W.}~\bibnamefont {Jang}}, \bibinfo {author} {\bibfnamefont {I.~M.}\ \bibnamefont {Shoemaker}}, \bibinfo {author} {\bibfnamefont {A.}~\bibnamefont {Thompson}}, \ and\ \bibinfo {author} {\bibfnamefont {J.}~\bibnamefont {Yu}},\ }\href {\doibase 10.1103/PhysRevD.104.115010} {\bibfield  {journal} {\bibinfo  {journal} {Phys. Rev. D}\ }\textbf {\bibinfo {volume} {104}},\ \bibinfo {pages} {115010} (\bibinfo {year} {2021})},\ \Eprint {http://arxiv.org/abs/2108.03262} {arXiv:2108.03262 [hep-ph]} \BibitemShut {NoStop}%
\bibitem [{\citenamefont {Andreev}\ \emph {et~al.}(2021)\citenamefont {Andreev} \emph {et~al.}}]{Andreev:2021fzd}%
  \BibitemOpen
  \bibfield  {author} {\bibinfo {author} {\bibfnamefont {Y.~M.}\ \bibnamefont {Andreev}} \emph {et~al.},\ }\href {\doibase 10.1103/PhysRevD.104.L091701} {\bibfield  {journal} {\bibinfo  {journal} {Phys. Rev. D}\ }\textbf {\bibinfo {volume} {104}},\ \bibinfo {pages} {L091701} (\bibinfo {year} {2021})},\ \Eprint {http://arxiv.org/abs/2108.04195} {arXiv:2108.04195 [hep-ex]} \BibitemShut {NoStop}%
\bibitem [{\citenamefont {Brdar}\ \emph {et~al.}(2023)\citenamefont {Brdar}, \citenamefont {Dutta}, \citenamefont {Jang}, \citenamefont {Kim}, \citenamefont {Shoemaker}, \citenamefont {Tabrizi}, \citenamefont {Thompson},\ and\ \citenamefont {Yu}}]{Brdar:2022vum}%
  \BibitemOpen
  \bibfield  {author} {\bibinfo {author} {\bibfnamefont {V.}~\bibnamefont {Brdar}}, \bibinfo {author} {\bibfnamefont {B.}~\bibnamefont {Dutta}}, \bibinfo {author} {\bibfnamefont {W.}~\bibnamefont {Jang}}, \bibinfo {author} {\bibfnamefont {D.}~\bibnamefont {Kim}}, \bibinfo {author} {\bibfnamefont {I.~M.}\ \bibnamefont {Shoemaker}}, \bibinfo {author} {\bibfnamefont {Z.}~\bibnamefont {Tabrizi}}, \bibinfo {author} {\bibfnamefont {A.}~\bibnamefont {Thompson}}, \ and\ \bibinfo {author} {\bibfnamefont {J.}~\bibnamefont {Yu}},\ }\href {\doibase 10.1103/PhysRevD.107.055043} {\bibfield  {journal} {\bibinfo  {journal} {Phys. Rev. D}\ }\textbf {\bibinfo {volume} {107}},\ \bibinfo {pages} {055043} (\bibinfo {year} {2023})},\ \Eprint {http://arxiv.org/abs/2206.06380} {arXiv:2206.06380 [hep-ph]} \BibitemShut {NoStop}%
\bibitem [{\citenamefont {Darm\'e}\ \emph {et~al.}(2022)\citenamefont {Darm\'e}, \citenamefont {Mancini}, \citenamefont {Nardi},\ and\ \citenamefont {Raggi}}]{Darme:2022zfw}%
  \BibitemOpen
  \bibfield  {author} {\bibinfo {author} {\bibfnamefont {L.}~\bibnamefont {Darm\'e}}, \bibinfo {author} {\bibfnamefont {M.}~\bibnamefont {Mancini}}, \bibinfo {author} {\bibfnamefont {E.}~\bibnamefont {Nardi}}, \ and\ \bibinfo {author} {\bibfnamefont {M.}~\bibnamefont {Raggi}},\ }\href {\doibase 10.1103/PhysRevD.106.115036} {\bibfield  {journal} {\bibinfo  {journal} {Phys. Rev. D}\ }\textbf {\bibinfo {volume} {106}},\ \bibinfo {pages} {115036} (\bibinfo {year} {2022})},\ \Eprint {http://arxiv.org/abs/2209.09261} {arXiv:2209.09261 [hep-ph]} \BibitemShut {NoStop}%
\bibitem [{\citenamefont {Andreev}\ \emph {et~al.}(2023)\citenamefont {Andreev} \emph {et~al.}}]{NA64:2023wbi}%
  \BibitemOpen
  \bibfield  {author} {\bibinfo {author} {\bibfnamefont {Y.~M.}\ \bibnamefont {Andreev}} \emph {et~al.} (\bibinfo {collaboration} {NA64}),\ }\href {\doibase 10.1103/PhysRevLett.131.161801} {\bibfield  {journal} {\bibinfo  {journal} {Phys. Rev. Lett.}\ }\textbf {\bibinfo {volume} {131}},\ \bibinfo {pages} {161801} (\bibinfo {year} {2023})},\ \Eprint {http://arxiv.org/abs/2307.02404} {arXiv:2307.02404 [hep-ex]} \BibitemShut {NoStop}%
\bibitem [{\citenamefont {Blinov}\ \emph {et~al.}(2024{\natexlab{a}})\citenamefont {Blinov}, \citenamefont {Gori},\ and\ \citenamefont {Hamer}}]{Blinov:2024gcw}%
  \BibitemOpen
  \bibfield  {author} {\bibinfo {author} {\bibfnamefont {N.}~\bibnamefont {Blinov}}, \bibinfo {author} {\bibfnamefont {S.}~\bibnamefont {Gori}}, \ and\ \bibinfo {author} {\bibfnamefont {N.}~\bibnamefont {Hamer}},\ }\href {\doibase 10.1103/PhysRevD.110.075006} {\bibfield  {journal} {\bibinfo  {journal} {Phys. Rev. D}\ }\textbf {\bibinfo {volume} {110}},\ \bibinfo {pages} {075006} (\bibinfo {year} {2024}{\natexlab{a}})},\ \Eprint {http://arxiv.org/abs/2405.17651} {arXiv:2405.17651 [hep-ph]} \BibitemShut {NoStop}%
\bibitem [{\citenamefont {Arias-Arag\'on}\ \emph {et~al.}(2024{\natexlab{a}})\citenamefont {Arias-Arag\'on}, \citenamefont {Darm\'e}, \citenamefont {di~Cortona},\ and\ \citenamefont {Nardi}}]{Arias-Aragon:2024qji}%
  \BibitemOpen
  \bibfield  {author} {\bibinfo {author} {\bibfnamefont {F.}~\bibnamefont {Arias-Arag\'on}}, \bibinfo {author} {\bibfnamefont {L.}~\bibnamefont {Darm\'e}}, \bibinfo {author} {\bibfnamefont {G.~G.}\ \bibnamefont {di~Cortona}}, \ and\ \bibinfo {author} {\bibfnamefont {E.}~\bibnamefont {Nardi}},\ }\href {\doibase 10.1103/PhysRevLett.132.261801} {\bibfield  {journal} {\bibinfo  {journal} {Phys. Rev. Lett.}\ }\textbf {\bibinfo {volume} {132}},\ \bibinfo {pages} {261801} (\bibinfo {year} {2024}{\natexlab{a}})},\ \Eprint {http://arxiv.org/abs/2403.15387} {arXiv:2403.15387 [hep-ph]} \BibitemShut {NoStop}%
\bibitem [{\citenamefont {Arias-Arag\'on}\ \emph {et~al.}(2024{\natexlab{b}})\citenamefont {Arias-Arag\'on}, \citenamefont {Darm\'e}, \citenamefont {di~Cortona},\ and\ \citenamefont {Nardi}}]{Arias-Aragon:2024gpm}%
  \BibitemOpen
  \bibfield  {author} {\bibinfo {author} {\bibfnamefont {F.}~\bibnamefont {Arias-Arag\'on}}, \bibinfo {author} {\bibfnamefont {L.}~\bibnamefont {Darm\'e}}, \bibinfo {author} {\bibfnamefont {G.~G.}\ \bibnamefont {di~Cortona}}, \ and\ \bibinfo {author} {\bibfnamefont {E.}~\bibnamefont {Nardi}},\ }\href@noop {} {\  (\bibinfo {year} {2024}{\natexlab{b}})},\ \Eprint {http://arxiv.org/abs/2407.15941} {arXiv:2407.15941 [hep-ph]} \BibitemShut {NoStop}%
\bibitem [{\citenamefont {Bauer}\ \emph {et~al.}(2018)\citenamefont {Bauer}, \citenamefont {Foldenauer},\ and\ \citenamefont {Jaeckel}}]{Bauer:2018onh}%
  \BibitemOpen
  \bibfield  {author} {\bibinfo {author} {\bibfnamefont {M.}~\bibnamefont {Bauer}}, \bibinfo {author} {\bibfnamefont {P.}~\bibnamefont {Foldenauer}}, \ and\ \bibinfo {author} {\bibfnamefont {J.}~\bibnamefont {Jaeckel}},\ }\href {\doibase 10.1007/JHEP07(2018)094} {\bibfield  {journal} {\bibinfo  {journal} {JHEP}\ }\textbf {\bibinfo {volume} {07}},\ \bibinfo {pages} {094} (\bibinfo {year} {2018})},\ \Eprint {http://arxiv.org/abs/1803.05466} {arXiv:1803.05466 [hep-ph]} \BibitemShut {NoStop}%
\bibitem [{\citenamefont {Ahdida}\ \emph {et~al.}(2021)\citenamefont {Ahdida} \emph {et~al.}}]{SHiP:2020vbd}%
  \BibitemOpen
  \bibfield  {author} {\bibinfo {author} {\bibfnamefont {C.}~\bibnamefont {Ahdida}} \emph {et~al.} (\bibinfo {collaboration} {SHiP}),\ }\href {\doibase 10.1140/epjc/s10052-021-09224-3} {\bibfield  {journal} {\bibinfo  {journal} {Eur. Phys. J. C}\ }\textbf {\bibinfo {volume} {81}},\ \bibinfo {pages} {451} (\bibinfo {year} {2021})},\ \Eprint {http://arxiv.org/abs/2011.05115} {arXiv:2011.05115 [hep-ex]} \BibitemShut {NoStop}%
\bibitem [{\citenamefont {Holdom}(1986)}]{Holdom:1985ag}%
  \BibitemOpen
  \bibfield  {author} {\bibinfo {author} {\bibfnamefont {B.}~\bibnamefont {Holdom}},\ }\href {\doibase 10.1016/0370-2693(86)91377-8} {\bibfield  {journal} {\bibinfo  {journal} {Phys. Lett. B}\ }\textbf {\bibinfo {volume} {166}},\ \bibinfo {pages} {196} (\bibinfo {year} {1986})}\BibitemShut {NoStop}%
\bibitem [{\citenamefont {Stueckelberg}(1938)}]{Stueckelberg:1938hvi}%
  \BibitemOpen
  \bibfield  {author} {\bibinfo {author} {\bibfnamefont {E.~C.~G.}\ \bibnamefont {Stueckelberg}},\ }\href {\doibase 10.5169/seals-110852} {\bibfield  {journal} {\bibinfo  {journal} {Helv. Phys. Acta}\ }\textbf {\bibinfo {volume} {11}},\ \bibinfo {pages} {225} (\bibinfo {year} {1938})}\BibitemShut {NoStop}%
\bibitem [{\citenamefont {He}\ \emph {et~al.}(1991)\citenamefont {He}, \citenamefont {Joshi}, \citenamefont {Lew},\ and\ \citenamefont {Volkas}}]{He:1991qd}%
  \BibitemOpen
  \bibfield  {author} {\bibinfo {author} {\bibfnamefont {X.-G.}\ \bibnamefont {He}}, \bibinfo {author} {\bibfnamefont {G.~C.}\ \bibnamefont {Joshi}}, \bibinfo {author} {\bibfnamefont {H.}~\bibnamefont {Lew}}, \ and\ \bibinfo {author} {\bibfnamefont {R.~R.}\ \bibnamefont {Volkas}},\ }\href {\doibase 10.1103/PhysRevD.44.2118} {\bibfield  {journal} {\bibinfo  {journal} {Phys. Rev. D}\ }\textbf {\bibinfo {volume} {44}},\ \bibinfo {pages} {2118} (\bibinfo {year} {1991})}\BibitemShut {NoStop}%
\bibitem [{\citenamefont {Araki}\ \emph {et~al.}(2012)\citenamefont {Araki}, \citenamefont {Heeck},\ and\ \citenamefont {Kubo}}]{Araki:2012ip}%
  \BibitemOpen
  \bibfield  {author} {\bibinfo {author} {\bibfnamefont {T.}~\bibnamefont {Araki}}, \bibinfo {author} {\bibfnamefont {J.}~\bibnamefont {Heeck}}, \ and\ \bibinfo {author} {\bibfnamefont {J.}~\bibnamefont {Kubo}},\ }\href {\doibase 10.1007/JHEP07(2012)083} {\bibfield  {journal} {\bibinfo  {journal} {JHEP}\ }\textbf {\bibinfo {volume} {07}},\ \bibinfo {pages} {083} (\bibinfo {year} {2012})},\ \Eprint {http://arxiv.org/abs/1203.4951} {arXiv:1203.4951 [hep-ph]} \BibitemShut {NoStop}%
\bibitem [{\citenamefont {Capozzi}\ \emph {et~al.}(2023)\citenamefont {Capozzi}, \citenamefont {Dutta}, \citenamefont {Gurung}, \citenamefont {Jang}, \citenamefont {Shoemaker}, \citenamefont {Thompson},\ and\ \citenamefont {Yu}}]{Capozzi:2023ffu}%
  \BibitemOpen
  \bibfield  {author} {\bibinfo {author} {\bibfnamefont {F.}~\bibnamefont {Capozzi}}, \bibinfo {author} {\bibfnamefont {B.}~\bibnamefont {Dutta}}, \bibinfo {author} {\bibfnamefont {G.}~\bibnamefont {Gurung}}, \bibinfo {author} {\bibfnamefont {W.}~\bibnamefont {Jang}}, \bibinfo {author} {\bibfnamefont {I.~M.}\ \bibnamefont {Shoemaker}}, \bibinfo {author} {\bibfnamefont {A.}~\bibnamefont {Thompson}}, \ and\ \bibinfo {author} {\bibfnamefont {J.}~\bibnamefont {Yu}},\ }\href {\doibase 10.1103/PhysRevD.108.075019} {\bibfield  {journal} {\bibinfo  {journal} {Phys. Rev. D}\ }\textbf {\bibinfo {volume} {108}},\ \bibinfo {pages} {075019} (\bibinfo {year} {2023})},\ \Eprint {http://arxiv.org/abs/2307.03878} {arXiv:2307.03878 [hep-ph]} \BibitemShut {NoStop}%
\bibitem [{\citenamefont {Lopez~Sola}\ \emph {et~al.}(2019)\citenamefont {Lopez~Sola} \emph {et~al.}}]{LopezSola:2019sfp}%
  \BibitemOpen
  \bibfield  {author} {\bibinfo {author} {\bibfnamefont {E.}~\bibnamefont {Lopez~Sola}} \emph {et~al.},\ }\href {\doibase 10.1103/PhysRevAccelBeams.22.113001} {\bibfield  {journal} {\bibinfo  {journal} {Phys. Rev. Accel. Beams}\ }\textbf {\bibinfo {volume} {22}},\ \bibinfo {pages} {113001} (\bibinfo {year} {2019})},\ \Eprint {http://arxiv.org/abs/1904.03074} {arXiv:1904.03074 [physics.ins-det]} \BibitemShut {NoStop}%
\bibitem [{\citenamefont {deNiverville}\ \emph {et~al.}(2011)\citenamefont {deNiverville}, \citenamefont {Pospelov},\ and\ \citenamefont {Ritz}}]{deNiverville:2011it}%
  \BibitemOpen
  \bibfield  {author} {\bibinfo {author} {\bibfnamefont {P.}~\bibnamefont {deNiverville}}, \bibinfo {author} {\bibfnamefont {M.}~\bibnamefont {Pospelov}}, \ and\ \bibinfo {author} {\bibfnamefont {A.}~\bibnamefont {Ritz}},\ }\href {\doibase 10.1103/PhysRevD.84.075020} {\bibfield  {journal} {\bibinfo  {journal} {Phys. Rev. D}\ }\textbf {\bibinfo {volume} {84}},\ \bibinfo {pages} {075020} (\bibinfo {year} {2011})},\ \Eprint {http://arxiv.org/abs/1107.4580} {arXiv:1107.4580 [hep-ph]} \BibitemShut {NoStop}%
\bibitem [{\citenamefont {Anelli}\ \emph {et~al.}(2015)\citenamefont {Anelli} \emph {et~al.}}]{SHiP:2015vad}%
  \BibitemOpen
  \bibfield  {author} {\bibinfo {author} {\bibfnamefont {M.}~\bibnamefont {Anelli}} \emph {et~al.} (\bibinfo {collaboration} {SHiP}),\ }\href@noop {} {\  (\bibinfo {year} {2015})},\ \Eprint {http://arxiv.org/abs/1504.04956} {arXiv:1504.04956 [physics.ins-det]} \BibitemShut {NoStop}%
\bibitem [{\citenamefont {Berryman}\ \emph {et~al.}(2020)\citenamefont {Berryman}, \citenamefont {de~Gouvea}, \citenamefont {Fox}, \citenamefont {Kayser}, \citenamefont {Kelly},\ and\ \citenamefont {Raaf}}]{Berryman:2019dme}%
  \BibitemOpen
  \bibfield  {author} {\bibinfo {author} {\bibfnamefont {J.~M.}\ \bibnamefont {Berryman}}, \bibinfo {author} {\bibfnamefont {A.}~\bibnamefont {de~Gouvea}}, \bibinfo {author} {\bibfnamefont {P.~J.}\ \bibnamefont {Fox}}, \bibinfo {author} {\bibfnamefont {B.~J.}\ \bibnamefont {Kayser}}, \bibinfo {author} {\bibfnamefont {K.~J.}\ \bibnamefont {Kelly}}, \ and\ \bibinfo {author} {\bibfnamefont {J.~L.}\ \bibnamefont {Raaf}},\ }\href {\doibase 10.1007/JHEP02(2020)174} {\bibfield  {journal} {\bibinfo  {journal} {JHEP}\ }\textbf {\bibinfo {volume} {02}},\ \bibinfo {pages} {174} (\bibinfo {year} {2020})},\ \Eprint {http://arxiv.org/abs/1912.07622} {arXiv:1912.07622 [hep-ph]} \BibitemShut {NoStop}%
\bibitem [{\citenamefont {Batell}\ \emph {et~al.}(2009)\citenamefont {Batell}, \citenamefont {Pospelov},\ and\ \citenamefont {Ritz}}]{Batell:2009di}%
  \BibitemOpen
  \bibfield  {author} {\bibinfo {author} {\bibfnamefont {B.}~\bibnamefont {Batell}}, \bibinfo {author} {\bibfnamefont {M.}~\bibnamefont {Pospelov}}, \ and\ \bibinfo {author} {\bibfnamefont {A.}~\bibnamefont {Ritz}},\ }\href {\doibase 10.1103/PhysRevD.80.095024} {\bibfield  {journal} {\bibinfo  {journal} {Phys. Rev. D}\ }\textbf {\bibinfo {volume} {80}},\ \bibinfo {pages} {095024} (\bibinfo {year} {2009})},\ \Eprint {http://arxiv.org/abs/0906.5614} {arXiv:0906.5614 [hep-ph]} \BibitemShut {NoStop}%
\bibitem [{\citenamefont {Essig}\ \emph {et~al.}(2010)\citenamefont {Essig}, \citenamefont {Harnik}, \citenamefont {Kaplan},\ and\ \citenamefont {Toro}}]{Essig:2010gu}%
  \BibitemOpen
  \bibfield  {author} {\bibinfo {author} {\bibfnamefont {R.}~\bibnamefont {Essig}}, \bibinfo {author} {\bibfnamefont {R.}~\bibnamefont {Harnik}}, \bibinfo {author} {\bibfnamefont {J.}~\bibnamefont {Kaplan}}, \ and\ \bibinfo {author} {\bibfnamefont {N.}~\bibnamefont {Toro}},\ }\href {\doibase 10.1103/PhysRevD.82.113008} {\bibfield  {journal} {\bibinfo  {journal} {Phys. Rev. D}\ }\textbf {\bibinfo {volume} {82}},\ \bibinfo {pages} {113008} (\bibinfo {year} {2010})},\ \Eprint {http://arxiv.org/abs/1008.0636} {arXiv:1008.0636 [hep-ph]} \BibitemShut {NoStop}%
\bibitem [{\citenamefont {deNiverville}\ \emph {et~al.}(2012)\citenamefont {deNiverville}, \citenamefont {McKeen},\ and\ \citenamefont {Ritz}}]{deNiverville:2012ij}%
  \BibitemOpen
  \bibfield  {author} {\bibinfo {author} {\bibfnamefont {P.}~\bibnamefont {deNiverville}}, \bibinfo {author} {\bibfnamefont {D.}~\bibnamefont {McKeen}}, \ and\ \bibinfo {author} {\bibfnamefont {A.}~\bibnamefont {Ritz}},\ }\href {\doibase 10.1103/PhysRevD.86.035022} {\bibfield  {journal} {\bibinfo  {journal} {Phys. Rev. D}\ }\textbf {\bibinfo {volume} {86}},\ \bibinfo {pages} {035022} (\bibinfo {year} {2012})},\ \Eprint {http://arxiv.org/abs/1205.3499} {arXiv:1205.3499 [hep-ph]} \BibitemShut {NoStop}%
\bibitem [{\citenamefont {Izaguirre}\ \emph {et~al.}(2013)\citenamefont {Izaguirre}, \citenamefont {Krnjaic}, \citenamefont {Schuster},\ and\ \citenamefont {Toro}}]{Izaguirre:2013uxa}%
  \BibitemOpen
  \bibfield  {author} {\bibinfo {author} {\bibfnamefont {E.}~\bibnamefont {Izaguirre}}, \bibinfo {author} {\bibfnamefont {G.}~\bibnamefont {Krnjaic}}, \bibinfo {author} {\bibfnamefont {P.}~\bibnamefont {Schuster}}, \ and\ \bibinfo {author} {\bibfnamefont {N.}~\bibnamefont {Toro}},\ }\href {\doibase 10.1103/PhysRevD.88.114015} {\bibfield  {journal} {\bibinfo  {journal} {Phys. Rev. D}\ }\textbf {\bibinfo {volume} {88}},\ \bibinfo {pages} {114015} (\bibinfo {year} {2013})},\ \Eprint {http://arxiv.org/abs/1307.6554} {arXiv:1307.6554 [hep-ph]} \BibitemShut {NoStop}%
\bibitem [{\citenamefont {Batell}\ \emph {et~al.}(2014)\citenamefont {Batell}, \citenamefont {deNiverville}, \citenamefont {McKeen}, \citenamefont {Pospelov},\ and\ \citenamefont {Ritz}}]{Batell:2014yra}%
  \BibitemOpen
  \bibfield  {author} {\bibinfo {author} {\bibfnamefont {B.}~\bibnamefont {Batell}}, \bibinfo {author} {\bibfnamefont {P.}~\bibnamefont {deNiverville}}, \bibinfo {author} {\bibfnamefont {D.}~\bibnamefont {McKeen}}, \bibinfo {author} {\bibfnamefont {M.}~\bibnamefont {Pospelov}}, \ and\ \bibinfo {author} {\bibfnamefont {A.}~\bibnamefont {Ritz}},\ }\href {\doibase 10.1103/PhysRevD.90.115014} {\bibfield  {journal} {\bibinfo  {journal} {Phys. Rev. D}\ }\textbf {\bibinfo {volume} {90}},\ \bibinfo {pages} {115014} (\bibinfo {year} {2014})},\ \Eprint {http://arxiv.org/abs/1405.7049} {arXiv:1405.7049 [hep-ph]} \BibitemShut {NoStop}%
\bibitem [{\citenamefont {Izaguirre}\ \emph {et~al.}(2015)\citenamefont {Izaguirre}, \citenamefont {Krnjaic}, \citenamefont {Schuster},\ and\ \citenamefont {Toro}}]{Izaguirre:2015yja}%
  \BibitemOpen
  \bibfield  {author} {\bibinfo {author} {\bibfnamefont {E.}~\bibnamefont {Izaguirre}}, \bibinfo {author} {\bibfnamefont {G.}~\bibnamefont {Krnjaic}}, \bibinfo {author} {\bibfnamefont {P.}~\bibnamefont {Schuster}}, \ and\ \bibinfo {author} {\bibfnamefont {N.}~\bibnamefont {Toro}},\ }\href {\doibase 10.1103/PhysRevLett.115.251301} {\bibfield  {journal} {\bibinfo  {journal} {Phys. Rev. Lett.}\ }\textbf {\bibinfo {volume} {115}},\ \bibinfo {pages} {251301} (\bibinfo {year} {2015})},\ \Eprint {http://arxiv.org/abs/1505.00011} {arXiv:1505.00011 [hep-ph]} \BibitemShut {NoStop}%
\bibitem [{\citenamefont {Coloma}\ \emph {et~al.}(2016)\citenamefont {Coloma}, \citenamefont {Dobrescu}, \citenamefont {Frugiuele},\ and\ \citenamefont {Harnik}}]{Coloma:2015pih}%
  \BibitemOpen
  \bibfield  {author} {\bibinfo {author} {\bibfnamefont {P.}~\bibnamefont {Coloma}}, \bibinfo {author} {\bibfnamefont {B.~A.}\ \bibnamefont {Dobrescu}}, \bibinfo {author} {\bibfnamefont {C.}~\bibnamefont {Frugiuele}}, \ and\ \bibinfo {author} {\bibfnamefont {R.}~\bibnamefont {Harnik}},\ }\href {\doibase 10.1007/JHEP04(2016)047} {\bibfield  {journal} {\bibinfo  {journal} {JHEP}\ }\textbf {\bibinfo {volume} {04}},\ \bibinfo {pages} {047} (\bibinfo {year} {2016})},\ \Eprint {http://arxiv.org/abs/1512.03852} {arXiv:1512.03852 [hep-ph]} \BibitemShut {NoStop}%
\bibitem [{\citenamefont {Alexander}\ \emph {et~al.}(2016)\citenamefont {Alexander} \emph {et~al.}}]{Alexander:2016aln}%
  \BibitemOpen
  \bibfield  {author} {\bibinfo {author} {\bibfnamefont {J.}~\bibnamefont {Alexander}} \emph {et~al.}\ }(\bibinfo {year} {2016})\ \Eprint {http://arxiv.org/abs/1608.08632} {arXiv:1608.08632 [hep-ph]} \BibitemShut {NoStop}%
\bibitem [{\citenamefont {deNiverville}\ \emph {et~al.}(2017)\citenamefont {deNiverville}, \citenamefont {Chen}, \citenamefont {Pospelov},\ and\ \citenamefont {Ritz}}]{deNiverville:2016rqh}%
  \BibitemOpen
  \bibfield  {author} {\bibinfo {author} {\bibfnamefont {P.}~\bibnamefont {deNiverville}}, \bibinfo {author} {\bibfnamefont {C.-Y.}\ \bibnamefont {Chen}}, \bibinfo {author} {\bibfnamefont {M.}~\bibnamefont {Pospelov}}, \ and\ \bibinfo {author} {\bibfnamefont {A.}~\bibnamefont {Ritz}},\ }\href {\doibase 10.1103/PhysRevD.95.035006} {\bibfield  {journal} {\bibinfo  {journal} {Phys. Rev. D}\ }\textbf {\bibinfo {volume} {95}},\ \bibinfo {pages} {035006} (\bibinfo {year} {2017})},\ \Eprint {http://arxiv.org/abs/1609.01770} {arXiv:1609.01770 [hep-ph]} \BibitemShut {NoStop}%
\bibitem [{\citenamefont {Magill}\ \emph {et~al.}(2018)\citenamefont {Magill}, \citenamefont {Plestid}, \citenamefont {Pospelov},\ and\ \citenamefont {Tsai}}]{Magill:2018jla}%
  \BibitemOpen
  \bibfield  {author} {\bibinfo {author} {\bibfnamefont {G.}~\bibnamefont {Magill}}, \bibinfo {author} {\bibfnamefont {R.}~\bibnamefont {Plestid}}, \bibinfo {author} {\bibfnamefont {M.}~\bibnamefont {Pospelov}}, \ and\ \bibinfo {author} {\bibfnamefont {Y.-D.}\ \bibnamefont {Tsai}},\ }\href {\doibase 10.1103/PhysRevD.98.115015} {\bibfield  {journal} {\bibinfo  {journal} {Phys. Rev. D}\ }\textbf {\bibinfo {volume} {98}},\ \bibinfo {pages} {115015} (\bibinfo {year} {2018})},\ \Eprint {http://arxiv.org/abs/1803.03262} {arXiv:1803.03262 [hep-ph]} \BibitemShut {NoStop}%
\bibitem [{\citenamefont {Berlin}\ \emph {et~al.}(2018)\citenamefont {Berlin}, \citenamefont {Gori}, \citenamefont {Schuster},\ and\ \citenamefont {Toro}}]{Berlin:2018pwi}%
  \BibitemOpen
  \bibfield  {author} {\bibinfo {author} {\bibfnamefont {A.}~\bibnamefont {Berlin}}, \bibinfo {author} {\bibfnamefont {S.}~\bibnamefont {Gori}}, \bibinfo {author} {\bibfnamefont {P.}~\bibnamefont {Schuster}}, \ and\ \bibinfo {author} {\bibfnamefont {N.}~\bibnamefont {Toro}},\ }\href {\doibase 10.1103/PhysRevD.98.035011} {\bibfield  {journal} {\bibinfo  {journal} {Phys. Rev. D}\ }\textbf {\bibinfo {volume} {98}},\ \bibinfo {pages} {035011} (\bibinfo {year} {2018})},\ \Eprint {http://arxiv.org/abs/1804.00661} {arXiv:1804.00661 [hep-ph]} \BibitemShut {NoStop}%
\bibitem [{\citenamefont {Aguilar-Arevalo}\ \emph {et~al.}(2018)\citenamefont {Aguilar-Arevalo} \emph {et~al.}}]{MiniBooNEDM:2018cxm}%
  \BibitemOpen
  \bibfield  {author} {\bibinfo {author} {\bibfnamefont {A.~A.}\ \bibnamefont {Aguilar-Arevalo}} \emph {et~al.} (\bibinfo {collaboration} {MiniBooNE DM}),\ }\href {\doibase 10.1103/PhysRevD.98.112004} {\bibfield  {journal} {\bibinfo  {journal} {Phys. Rev. D}\ }\textbf {\bibinfo {volume} {98}},\ \bibinfo {pages} {112004} (\bibinfo {year} {2018})},\ \Eprint {http://arxiv.org/abs/1807.06137} {arXiv:1807.06137 [hep-ex]} \BibitemShut {NoStop}%
\bibitem [{\citenamefont {Magill}\ \emph {et~al.}(2019)\citenamefont {Magill}, \citenamefont {Plestid}, \citenamefont {Pospelov},\ and\ \citenamefont {Tsai}}]{Magill:2018tbb}%
  \BibitemOpen
  \bibfield  {author} {\bibinfo {author} {\bibfnamefont {G.}~\bibnamefont {Magill}}, \bibinfo {author} {\bibfnamefont {R.}~\bibnamefont {Plestid}}, \bibinfo {author} {\bibfnamefont {M.}~\bibnamefont {Pospelov}}, \ and\ \bibinfo {author} {\bibfnamefont {Y.-D.}\ \bibnamefont {Tsai}},\ }\href {\doibase 10.1103/PhysRevLett.122.071801} {\bibfield  {journal} {\bibinfo  {journal} {Phys. Rev. Lett.}\ }\textbf {\bibinfo {volume} {122}},\ \bibinfo {pages} {071801} (\bibinfo {year} {2019})},\ \Eprint {http://arxiv.org/abs/1806.03310} {arXiv:1806.03310 [hep-ph]} \BibitemShut {NoStop}%
\bibitem [{\citenamefont {Plestid}\ \emph {et~al.}(2020)\citenamefont {Plestid}, \citenamefont {Takhistov}, \citenamefont {Tsai}, \citenamefont {Bringmann}, \citenamefont {Kusenko},\ and\ \citenamefont {Pospelov}}]{Plestid:2020kdm}%
  \BibitemOpen
  \bibfield  {author} {\bibinfo {author} {\bibfnamefont {R.}~\bibnamefont {Plestid}}, \bibinfo {author} {\bibfnamefont {V.}~\bibnamefont {Takhistov}}, \bibinfo {author} {\bibfnamefont {Y.-D.}\ \bibnamefont {Tsai}}, \bibinfo {author} {\bibfnamefont {T.}~\bibnamefont {Bringmann}}, \bibinfo {author} {\bibfnamefont {A.}~\bibnamefont {Kusenko}}, \ and\ \bibinfo {author} {\bibfnamefont {M.}~\bibnamefont {Pospelov}},\ }\href {\doibase 10.1103/PhysRevD.102.115032} {\bibfield  {journal} {\bibinfo  {journal} {Phys. Rev. D}\ }\textbf {\bibinfo {volume} {102}},\ \bibinfo {pages} {115032} (\bibinfo {year} {2020})},\ \Eprint {http://arxiv.org/abs/2002.11732} {arXiv:2002.11732 [hep-ph]} \BibitemShut {NoStop}%
\bibitem [{\citenamefont {Berlin}\ \emph {et~al.}(2020)\citenamefont {Berlin}, \citenamefont {deNiverville}, \citenamefont {Ritz}, \citenamefont {Schuster},\ and\ \citenamefont {Toro}}]{Berlin:2020uwy}%
  \BibitemOpen
  \bibfield  {author} {\bibinfo {author} {\bibfnamefont {A.}~\bibnamefont {Berlin}}, \bibinfo {author} {\bibfnamefont {P.}~\bibnamefont {deNiverville}}, \bibinfo {author} {\bibfnamefont {A.}~\bibnamefont {Ritz}}, \bibinfo {author} {\bibfnamefont {P.}~\bibnamefont {Schuster}}, \ and\ \bibinfo {author} {\bibfnamefont {N.}~\bibnamefont {Toro}},\ }\href {\doibase 10.1103/PhysRevD.102.095011} {\bibfield  {journal} {\bibinfo  {journal} {Phys. Rev. D}\ }\textbf {\bibinfo {volume} {102}},\ \bibinfo {pages} {095011} (\bibinfo {year} {2020})},\ \Eprint {http://arxiv.org/abs/2003.03379} {arXiv:2003.03379 [hep-ph]} \BibitemShut {NoStop}%
\bibitem [{\citenamefont {Batell}\ \emph {et~al.}(2021)\citenamefont {Batell}, \citenamefont {Evans}, \citenamefont {Gori},\ and\ \citenamefont {Rai}}]{Batell:2020vqn}%
  \BibitemOpen
  \bibfield  {author} {\bibinfo {author} {\bibfnamefont {B.}~\bibnamefont {Batell}}, \bibinfo {author} {\bibfnamefont {J.~A.}\ \bibnamefont {Evans}}, \bibinfo {author} {\bibfnamefont {S.}~\bibnamefont {Gori}}, \ and\ \bibinfo {author} {\bibfnamefont {M.}~\bibnamefont {Rai}},\ }\href {\doibase 10.1007/JHEP05(2021)049} {\bibfield  {journal} {\bibinfo  {journal} {JHEP}\ }\textbf {\bibinfo {volume} {05}},\ \bibinfo {pages} {049} (\bibinfo {year} {2021})},\ \Eprint {http://arxiv.org/abs/2008.08108} {arXiv:2008.08108 [hep-ph]} \BibitemShut {NoStop}%
\bibitem [{\citenamefont {De~Romeri}\ \emph {et~al.}(2019)\citenamefont {De~Romeri}, \citenamefont {Kelly},\ and\ \citenamefont {Machado}}]{DeRomeri:2019kic}%
  \BibitemOpen
  \bibfield  {author} {\bibinfo {author} {\bibfnamefont {V.}~\bibnamefont {De~Romeri}}, \bibinfo {author} {\bibfnamefont {K.~J.}\ \bibnamefont {Kelly}}, \ and\ \bibinfo {author} {\bibfnamefont {P.~A.~N.}\ \bibnamefont {Machado}},\ }\href {\doibase 10.1103/PhysRevD.100.095010} {\bibfield  {journal} {\bibinfo  {journal} {Phys. Rev. D}\ }\textbf {\bibinfo {volume} {100}},\ \bibinfo {pages} {095010} (\bibinfo {year} {2019})},\ \Eprint {http://arxiv.org/abs/1903.10505} {arXiv:1903.10505 [hep-ph]} \BibitemShut {NoStop}%
\bibitem [{\citenamefont {Kelly}\ and\ \citenamefont {Machado}(2021)}]{Kelly:2021xbv}%
  \BibitemOpen
  \bibfield  {author} {\bibinfo {author} {\bibfnamefont {K.~J.}\ \bibnamefont {Kelly}}\ and\ \bibinfo {author} {\bibfnamefont {P.~A.~N.}\ \bibnamefont {Machado}},\ }\href {\doibase 10.1103/PhysRevD.104.055015} {\bibfield  {journal} {\bibinfo  {journal} {Phys. Rev. D}\ }\textbf {\bibinfo {volume} {104}},\ \bibinfo {pages} {055015} (\bibinfo {year} {2021})},\ \Eprint {http://arxiv.org/abs/2106.06548} {arXiv:2106.06548 [hep-ph]} \BibitemShut {NoStop}%
\bibitem [{\citenamefont {Breitbach}\ \emph {et~al.}(2022)\citenamefont {Breitbach}, \citenamefont {Buonocore}, \citenamefont {Frugiuele}, \citenamefont {Kopp},\ and\ \citenamefont {Mittnacht}}]{Breitbach:2021gvv}%
  \BibitemOpen
  \bibfield  {author} {\bibinfo {author} {\bibfnamefont {M.}~\bibnamefont {Breitbach}}, \bibinfo {author} {\bibfnamefont {L.}~\bibnamefont {Buonocore}}, \bibinfo {author} {\bibfnamefont {C.}~\bibnamefont {Frugiuele}}, \bibinfo {author} {\bibfnamefont {J.}~\bibnamefont {Kopp}}, \ and\ \bibinfo {author} {\bibfnamefont {L.}~\bibnamefont {Mittnacht}},\ }\href {\doibase 10.1007/JHEP01(2022)048} {\bibfield  {journal} {\bibinfo  {journal} {JHEP}\ }\textbf {\bibinfo {volume} {01}},\ \bibinfo {pages} {048} (\bibinfo {year} {2022})},\ \Eprint {http://arxiv.org/abs/2102.03383} {arXiv:2102.03383 [hep-ph]} \BibitemShut {NoStop}%
\bibitem [{\citenamefont {Dev}\ \emph {et~al.}(2021)\citenamefont {Dev}, \citenamefont {Dutta}, \citenamefont {Kelly}, \citenamefont {Mohapatra},\ and\ \citenamefont {Zhang}}]{Dev:2021qjj}%
  \BibitemOpen
  \bibfield  {author} {\bibinfo {author} {\bibfnamefont {P.~S.~B.}\ \bibnamefont {Dev}}, \bibinfo {author} {\bibfnamefont {B.}~\bibnamefont {Dutta}}, \bibinfo {author} {\bibfnamefont {K.~J.}\ \bibnamefont {Kelly}}, \bibinfo {author} {\bibfnamefont {R.~N.}\ \bibnamefont {Mohapatra}}, \ and\ \bibinfo {author} {\bibfnamefont {Y.}~\bibnamefont {Zhang}},\ }\href {\doibase 10.1007/JHEP07(2021)166} {\bibfield  {journal} {\bibinfo  {journal} {JHEP}\ }\textbf {\bibinfo {volume} {07}},\ \bibinfo {pages} {166} (\bibinfo {year} {2021})},\ \Eprint {http://arxiv.org/abs/2104.07681} {arXiv:2104.07681 [hep-ph]} \BibitemShut {NoStop}%
\bibitem [{\citenamefont {Blinov}\ \emph {et~al.}(2022)\citenamefont {Blinov}, \citenamefont {Kowalczyk},\ and\ \citenamefont {Wynne}}]{Blinov:2021say}%
  \BibitemOpen
  \bibfield  {author} {\bibinfo {author} {\bibfnamefont {N.}~\bibnamefont {Blinov}}, \bibinfo {author} {\bibfnamefont {E.}~\bibnamefont {Kowalczyk}}, \ and\ \bibinfo {author} {\bibfnamefont {M.}~\bibnamefont {Wynne}},\ }\href {\doibase 10.1007/JHEP02(2022)036} {\bibfield  {journal} {\bibinfo  {journal} {JHEP}\ }\textbf {\bibinfo {volume} {02}},\ \bibinfo {pages} {036} (\bibinfo {year} {2022})},\ \Eprint {http://arxiv.org/abs/2112.09814} {arXiv:2112.09814 [hep-ph]} \BibitemShut {NoStop}%
\bibitem [{\citenamefont {Gori}\ \emph {et~al.}(2022)\citenamefont {Gori} \emph {et~al.}}]{Gori:2022vri}%
  \BibitemOpen
  \bibfield  {author} {\bibinfo {author} {\bibfnamefont {S.}~\bibnamefont {Gori}} \emph {et~al.},\ }\href@noop {} {\  (\bibinfo {year} {2022})},\ \Eprint {http://arxiv.org/abs/2209.04671} {arXiv:2209.04671 [hep-ph]} \BibitemShut {NoStop}%
\bibitem [{\citenamefont {Coloma}\ \emph {et~al.}(2024)\citenamefont {Coloma}, \citenamefont {L\'opez-Pav\'on}, \citenamefont {Molina-Bueno},\ and\ \citenamefont {Urrea}}]{Coloma:2023adi}%
  \BibitemOpen
  \bibfield  {author} {\bibinfo {author} {\bibfnamefont {P.}~\bibnamefont {Coloma}}, \bibinfo {author} {\bibfnamefont {J.}~\bibnamefont {L\'opez-Pav\'on}}, \bibinfo {author} {\bibfnamefont {L.}~\bibnamefont {Molina-Bueno}}, \ and\ \bibinfo {author} {\bibfnamefont {S.}~\bibnamefont {Urrea}},\ }\href {\doibase 10.1007/JHEP01(2024)134} {\bibfield  {journal} {\bibinfo  {journal} {JHEP}\ }\textbf {\bibinfo {volume} {01}},\ \bibinfo {pages} {134} (\bibinfo {year} {2024})},\ \Eprint {http://arxiv.org/abs/2304.06765} {arXiv:2304.06765 [hep-ph]} \BibitemShut {NoStop}%
\bibitem [{\citenamefont {Blinov}\ \emph {et~al.}(2024{\natexlab{b}})\citenamefont {Blinov}, \citenamefont {Fox}, \citenamefont {Kelly}, \citenamefont {Machado},\ and\ \citenamefont {Plestid}}]{Blinov:2024pza}%
  \BibitemOpen
  \bibfield  {author} {\bibinfo {author} {\bibfnamefont {N.}~\bibnamefont {Blinov}}, \bibinfo {author} {\bibfnamefont {P.~J.}\ \bibnamefont {Fox}}, \bibinfo {author} {\bibfnamefont {K.~J.}\ \bibnamefont {Kelly}}, \bibinfo {author} {\bibfnamefont {P.~A.~N.}\ \bibnamefont {Machado}}, \ and\ \bibinfo {author} {\bibfnamefont {R.}~\bibnamefont {Plestid}},\ }\href {\doibase 10.1007/JHEP07(2024)022} {\bibfield  {journal} {\bibinfo  {journal} {JHEP}\ }\textbf {\bibinfo {volume} {07}},\ \bibinfo {pages} {022} (\bibinfo {year} {2024}{\natexlab{b}})},\ \Eprint {http://arxiv.org/abs/2401.06843} {arXiv:2401.06843 [hep-ph]} \BibitemShut {NoStop}%
\bibitem [{\citenamefont {Bierlich}\ \emph {et~al.}(2022)\citenamefont {Bierlich} \emph {et~al.}}]{Bierlich:2022pfr}%
  \BibitemOpen
  \bibfield  {author} {\bibinfo {author} {\bibfnamefont {C.}~\bibnamefont {Bierlich}} \emph {et~al.},\ }\href {\doibase 10.21468/SciPostPhysCodeb.8} {\bibfield  {journal} {\bibinfo  {journal} {SciPost Phys. Codeb.}\ }\textbf {\bibinfo {volume} {2022}},\ \bibinfo {pages} {8} (\bibinfo {year} {2022})},\ \Eprint {http://arxiv.org/abs/2203.11601} {arXiv:2203.11601 [hep-ph]} \BibitemShut {NoStop}%
\bibitem [{\citenamefont {Plestid}\ and\ \citenamefont {Wise}(2024{\natexlab{a}})}]{Plestid:2024jqm}%
  \BibitemOpen
  \bibfield  {author} {\bibinfo {author} {\bibfnamefont {R.}~\bibnamefont {Plestid}}\ and\ \bibinfo {author} {\bibfnamefont {M.~B.}\ \bibnamefont {Wise}},\ }\href@noop {} {\  (\bibinfo {year} {2024}{\natexlab{a}})},\ \Eprint {http://arxiv.org/abs/2407.21752} {arXiv:2407.21752 [hep-ph]} \BibitemShut {NoStop}%
\bibitem [{\citenamefont {Plestid}\ and\ \citenamefont {Wise}(2024{\natexlab{b}})}]{Plestid:2024xzh}%
  \BibitemOpen
  \bibfield  {author} {\bibinfo {author} {\bibfnamefont {R.}~\bibnamefont {Plestid}}\ and\ \bibinfo {author} {\bibfnamefont {M.~B.}\ \bibnamefont {Wise}},\ }\href {\doibase 10.1103/PhysRevD.110.056032} {\bibfield  {journal} {\bibinfo  {journal} {Phys. Rev. D}\ }\textbf {\bibinfo {volume} {110}},\ \bibinfo {pages} {056032} (\bibinfo {year} {2024}{\natexlab{b}})},\ \Eprint {http://arxiv.org/abs/2403.12184} {arXiv:2403.12184 [hep-ph]} \BibitemShut {NoStop}%
\bibitem [{\citenamefont {Slater}(1930)}]{Slater:1930APS}%
  \BibitemOpen
  \bibfield  {author} {\bibinfo {author} {\bibfnamefont {J.~C.}\ \bibnamefont {Slater}},\ }\href {\doibase 10.1103/PhysRev.36.57} {\bibfield  {journal} {\bibinfo  {journal} {Phys. Rev.}\ }\textbf {\bibinfo {volume} {36}},\ \bibinfo {pages} {57} (\bibinfo {year} {1930})}\BibitemShut {NoStop}%
\bibitem [{\citenamefont {Machado Santos~Soares}(2021)}]{MachadoSantosSoares:2765979}%
  \BibitemOpen
  \bibfield  {author} {\bibinfo {author} {\bibfnamefont {G.}~\bibnamefont {Machado Santos~Soares}},\ }\emph {\bibinfo {title} {{Optimization of the Selection of Hidden Particles in the SHiP Experiment}}},\ \href {https://cds.cern.ch/record/2765979} {Master's thesis},\ \bibinfo  {school} {Instituto Superior Técnico - University of Lisbon} (\bibinfo {year} {2021}),\ \bibinfo {note} {presented 26 Jan 2021}\BibitemShut {NoStop}%
\bibitem [{\citenamefont {Serra}(2024)}]{Nico:chat}%
  \BibitemOpen
  \bibfield  {author} {\bibinfo {author} {\bibfnamefont {N.}~\bibnamefont {Serra}},\ }\href@noop {} {} (\bibinfo {year} {2024}),\ \bibinfo {note} {private communication}\BibitemShut {NoStop}%
\bibitem [{\citenamefont {Chang}\ \emph {et~al.}(2017)\citenamefont {Chang}, \citenamefont {Essig},\ and\ \citenamefont {McDermott}}]{Chang:2016ntp}%
  \BibitemOpen
  \bibfield  {author} {\bibinfo {author} {\bibfnamefont {J.~H.}\ \bibnamefont {Chang}}, \bibinfo {author} {\bibfnamefont {R.}~\bibnamefont {Essig}}, \ and\ \bibinfo {author} {\bibfnamefont {S.~D.}\ \bibnamefont {McDermott}},\ }\href {\doibase 10.1007/JHEP01(2017)107} {\bibfield  {journal} {\bibinfo  {journal} {JHEP}\ }\textbf {\bibinfo {volume} {01}},\ \bibinfo {pages} {107} (\bibinfo {year} {2017})},\ \Eprint {http://arxiv.org/abs/1611.03864} {arXiv:1611.03864 [hep-ph]} \BibitemShut {NoStop}%
\bibitem [{\citenamefont {Batley}\ \emph {et~al.}(2015)\citenamefont {Batley} \emph {et~al.}}]{NA482:2015wmo}%
  \BibitemOpen
  \bibfield  {author} {\bibinfo {author} {\bibfnamefont {J.~R.}\ \bibnamefont {Batley}} \emph {et~al.} (\bibinfo {collaboration} {NA48/2}),\ }\href {\doibase 10.1016/j.physletb.2015.04.068} {\bibfield  {journal} {\bibinfo  {journal} {Phys. Lett. B}\ }\textbf {\bibinfo {volume} {746}},\ \bibinfo {pages} {178} (\bibinfo {year} {2015})},\ \Eprint {http://arxiv.org/abs/1504.00607} {arXiv:1504.00607 [hep-ex]} \BibitemShut {NoStop}%
\bibitem [{\citenamefont {Lees}\ \emph {et~al.}(2014)\citenamefont {Lees} \emph {et~al.}}]{BaBar:2014zli}%
  \BibitemOpen
  \bibfield  {author} {\bibinfo {author} {\bibfnamefont {J.~P.}\ \bibnamefont {Lees}} \emph {et~al.} (\bibinfo {collaboration} {BaBar}),\ }\href {\doibase 10.1103/PhysRevLett.113.201801} {\bibfield  {journal} {\bibinfo  {journal} {Phys. Rev. Lett.}\ }\textbf {\bibinfo {volume} {113}},\ \bibinfo {pages} {201801} (\bibinfo {year} {2014})},\ \Eprint {http://arxiv.org/abs/1406.2980} {arXiv:1406.2980 [hep-ex]} \BibitemShut {NoStop}%
\bibitem [{\citenamefont {Lees}\ \emph {et~al.}(2016)\citenamefont {Lees} \emph {et~al.}}]{BaBar:2016sci}%
  \BibitemOpen
  \bibfield  {author} {\bibinfo {author} {\bibfnamefont {J.~P.}\ \bibnamefont {Lees}} \emph {et~al.} (\bibinfo {collaboration} {BaBar}),\ }\href {\doibase 10.1103/PhysRevD.94.011102} {\bibfield  {journal} {\bibinfo  {journal} {Phys. Rev. D}\ }\textbf {\bibinfo {volume} {94}},\ \bibinfo {pages} {011102} (\bibinfo {year} {2016})},\ \Eprint {http://arxiv.org/abs/1606.03501} {arXiv:1606.03501 [hep-ex]} \BibitemShut {NoStop}%
\bibitem [{\citenamefont {Aaij}\ \emph {et~al.}(2020)\citenamefont {Aaij} \emph {et~al.}}]{LHCb:2019vmc}%
  \BibitemOpen
  \bibfield  {author} {\bibinfo {author} {\bibfnamefont {R.}~\bibnamefont {Aaij}} \emph {et~al.} (\bibinfo {collaboration} {LHCb}),\ }\href {\doibase 10.1103/PhysRevLett.124.041801} {\bibfield  {journal} {\bibinfo  {journal} {Phys. Rev. Lett.}\ }\textbf {\bibinfo {volume} {124}},\ \bibinfo {pages} {041801} (\bibinfo {year} {2020})},\ \Eprint {http://arxiv.org/abs/1910.06926} {arXiv:1910.06926 [hep-ex]} \BibitemShut {NoStop}%
\bibitem [{\citenamefont {Ilten}\ \emph {et~al.}(2018)\citenamefont {Ilten}, \citenamefont {Soreq}, \citenamefont {Williams},\ and\ \citenamefont {Xue}}]{Ilten:2018crw}%
  \BibitemOpen
  \bibfield  {author} {\bibinfo {author} {\bibfnamefont {P.}~\bibnamefont {Ilten}}, \bibinfo {author} {\bibfnamefont {Y.}~\bibnamefont {Soreq}}, \bibinfo {author} {\bibfnamefont {M.}~\bibnamefont {Williams}}, \ and\ \bibinfo {author} {\bibfnamefont {W.}~\bibnamefont {Xue}},\ }\href {\doibase 10.1007/JHEP06(2018)004} {\bibfield  {journal} {\bibinfo  {journal} {JHEP}\ }\textbf {\bibinfo {volume} {06}},\ \bibinfo {pages} {004} (\bibinfo {year} {2018})},\ \Eprint {http://arxiv.org/abs/1801.04847} {arXiv:1801.04847 [hep-ph]} \BibitemShut {NoStop}%
\bibitem [{\citenamefont {Baruch}\ \emph {et~al.}(2022)\citenamefont {Baruch}, \citenamefont {Ilten}, \citenamefont {Soreq},\ and\ \citenamefont {Williams}}]{Baruch:2022esd}%
  \BibitemOpen
  \bibfield  {author} {\bibinfo {author} {\bibfnamefont {C.}~\bibnamefont {Baruch}}, \bibinfo {author} {\bibfnamefont {P.}~\bibnamefont {Ilten}}, \bibinfo {author} {\bibfnamefont {Y.}~\bibnamefont {Soreq}}, \ and\ \bibinfo {author} {\bibfnamefont {M.}~\bibnamefont {Williams}},\ }\href {\doibase 10.1007/JHEP11(2022)124} {\bibfield  {journal} {\bibinfo  {journal} {JHEP}\ }\textbf {\bibinfo {volume} {11}},\ \bibinfo {pages} {124} (\bibinfo {year} {2022})},\ \Eprint {http://arxiv.org/abs/2206.08563} {arXiv:2206.08563 [hep-ph]} \BibitemShut {NoStop}%
\bibitem [{\citenamefont {Deniz}\ \emph {et~al.}(2010)\citenamefont {Deniz} \emph {et~al.}}]{TEXONO:2009knm}%
  \BibitemOpen
  \bibfield  {author} {\bibinfo {author} {\bibfnamefont {M.}~\bibnamefont {Deniz}} \emph {et~al.} (\bibinfo {collaboration} {TEXONO}),\ }\href {\doibase 10.1103/PhysRevD.81.072001} {\bibfield  {journal} {\bibinfo  {journal} {Phys. Rev. D}\ }\textbf {\bibinfo {volume} {81}},\ \bibinfo {pages} {072001} (\bibinfo {year} {2010})},\ \Eprint {http://arxiv.org/abs/0911.1597} {arXiv:0911.1597 [hep-ex]} \BibitemShut {NoStop}%
\bibitem [{\citenamefont {Bilmis}\ \emph {et~al.}(2015)\citenamefont {Bilmis}, \citenamefont {Turan}, \citenamefont {Aliev}, \citenamefont {Deniz}, \citenamefont {Singh},\ and\ \citenamefont {Wong}}]{Bilmis:2015lja}%
  \BibitemOpen
  \bibfield  {author} {\bibinfo {author} {\bibfnamefont {S.}~\bibnamefont {Bilmis}}, \bibinfo {author} {\bibfnamefont {I.}~\bibnamefont {Turan}}, \bibinfo {author} {\bibfnamefont {T.~M.}\ \bibnamefont {Aliev}}, \bibinfo {author} {\bibfnamefont {M.}~\bibnamefont {Deniz}}, \bibinfo {author} {\bibfnamefont {L.}~\bibnamefont {Singh}}, \ and\ \bibinfo {author} {\bibfnamefont {H.~T.}\ \bibnamefont {Wong}},\ }\href {\doibase 10.1103/PhysRevD.92.033009} {\bibfield  {journal} {\bibinfo  {journal} {Phys. Rev. D}\ }\textbf {\bibinfo {volume} {92}},\ \bibinfo {pages} {033009} (\bibinfo {year} {2015})},\ \Eprint {http://arxiv.org/abs/1502.07763} {arXiv:1502.07763 [hep-ph]} \BibitemShut {NoStop}%
\bibitem [{\citenamefont {Lindner}\ \emph {et~al.}(2018)\citenamefont {Lindner}, \citenamefont {Queiroz}, \citenamefont {Rodejohann},\ and\ \citenamefont {Xu}}]{Lindner:2018kjo}%
  \BibitemOpen
  \bibfield  {author} {\bibinfo {author} {\bibfnamefont {M.}~\bibnamefont {Lindner}}, \bibinfo {author} {\bibfnamefont {F.~S.}\ \bibnamefont {Queiroz}}, \bibinfo {author} {\bibfnamefont {W.}~\bibnamefont {Rodejohann}}, \ and\ \bibinfo {author} {\bibfnamefont {X.-J.}\ \bibnamefont {Xu}},\ }\href {\doibase 10.1007/JHEP05(2018)098} {\bibfield  {journal} {\bibinfo  {journal} {JHEP}\ }\textbf {\bibinfo {volume} {05}},\ \bibinfo {pages} {098} (\bibinfo {year} {2018})},\ \Eprint {http://arxiv.org/abs/1803.00060} {arXiv:1803.00060 [hep-ph]} \BibitemShut {NoStop}%
\bibitem [{\citenamefont {Wise}\ and\ \citenamefont {Zhang}(2018)}]{Wise:2018rnb}%
  \BibitemOpen
  \bibfield  {author} {\bibinfo {author} {\bibfnamefont {M.~B.}\ \bibnamefont {Wise}}\ and\ \bibinfo {author} {\bibfnamefont {Y.}~\bibnamefont {Zhang}},\ }\href {\doibase 10.1007/JHEP06(2018)053} {\bibfield  {journal} {\bibinfo  {journal} {JHEP}\ }\textbf {\bibinfo {volume} {06}},\ \bibinfo {pages} {053} (\bibinfo {year} {2018})},\ \Eprint {http://arxiv.org/abs/1803.00591} {arXiv:1803.00591 [hep-ph]} \BibitemShut {NoStop}%
\bibitem [{\citenamefont {Kuraev}\ and\ \citenamefont {Fadin}(1985)}]{Kuraev:1985hb}%
  \BibitemOpen
  \bibfield  {author} {\bibinfo {author} {\bibfnamefont {E.~A.}\ \bibnamefont {Kuraev}}\ and\ \bibinfo {author} {\bibfnamefont {V.~S.}\ \bibnamefont {Fadin}},\ }\href@noop {} {\bibfield  {journal} {\bibinfo  {journal} {Sov. J. Nucl. Phys.}\ }\textbf {\bibinfo {volume} {41}},\ \bibinfo {pages} {466} (\bibinfo {year} {1985})}\BibitemShut {NoStop}%
\bibitem [{\citenamefont {Nicrosini}\ and\ \citenamefont {Trentadue}(1987)}]{Nicrosini:1986sm}%
  \BibitemOpen
  \bibfield  {author} {\bibinfo {author} {\bibfnamefont {O.}~\bibnamefont {Nicrosini}}\ and\ \bibinfo {author} {\bibfnamefont {L.}~\bibnamefont {Trentadue}},\ }\href {\doibase 10.1016/0370-2693(87)90819-7} {\bibfield  {journal} {\bibinfo  {journal} {Phys. Lett. B}\ }\textbf {\bibinfo {volume} {196}},\ \bibinfo {pages} {551} (\bibinfo {year} {1987})}\BibitemShut {NoStop}%
\end{thebibliography}%

\end{document}